\documentclass[12pt]{article}
\begin{document}
      \sloppy
      %\large
%%%%%%%%%%%%%%%%%%%%%%%%%%%%%%%%%%%%%%%%%%%%%%%%%%%%%%%%%%%%%%%%%%

%%%%%%%%% This defines A4 output %%%%%%%%%%%%%%%%%%%%%%%%%%%
\def\AFOUR{%
\setlength{\textheight}{9.0in}%
\setlength{\textwidth}{5.75in}%
\setlength{\topmargin}{-0.375in}%
\hoffset=-.5in%
\renewcommand{\baselinestretch}{1.17}%
\setlength{\parskip}{6pt plus 2pt}%
}
%%% Uncomment the following line to obtain A4 output %%%%%%%
\AFOUR
%%%%%%%%%%%%%%%%%%%%%%%%%%%%%%%%%%%%%%%%%%%%%%%%%%%%%%%
%%%
\def\car{\mathop{\square}}
\def\carre#1#2{\raise 2pt\hbox{$\scriptstyle #1$}\car_{#2}}

\parindent=0pt
%%%%%%%% This deflates (sub)section titles %%%%%%%%%%%%%%
\makeatletter
\def\section{\@startsection {section}{1}{\z@}{-3.5ex plus -1ex minus
   -.2ex}{2.3ex plus .2ex}{\large\bf}}
\def\subsection{\@startsection{subsection}{2}{\z@}{-3.25ex plus -1ex minus
   -.2ex}{1.5ex plus .2ex}{\normalsize\bf}}
\makeatother
%%%%%%%%% This numbers equations by sections %%%%%%%%%%%%%
\makeatletter
\@addtoreset{equation}{section}
\renewcommand{\theequation}{\thesection.\arabic{equation}}
\makeatother

%%%%%%%%% Nicola %%%%%%%%%%%%%%%%%%%%%%%%%%%%%%%%%%%%%%%%
%LETTRES GRECQUES
\renewcommand{\a}{\alpha}
\renewcommand{\b}{\beta}
\newcommand{\g}{\gamma}           \newcommand{\G}{\Gamma}
\renewcommand{\d}{\delta}         \newcommand{\D}{\Delta}
\newcommand{\e}{\varepsilon}
\newcommand{\la}{\lambda}        \newcommand{\LA}{\Lambda}
\newcommand{\m}{\mu}
\newcommand{\A}{\widehat{A}^{\star a}_{\mu}}
\newcommand{\Ar}{\widehat{A}^{\star a}_{\rho}}
\newcommand{\n}{\nu}
\newcommand{\om}{\omega}         \newcommand{\OM}{\Omega}
\newcommand{\p}{\psi}             \newcommand{\PS}{\Psi}
\renewcommand{\r}{\rho}
\newcommand{\s}{\sigma}           \renewcommand{\S}{\Sigma}
\newcommand{\vf}{\varphi}
\newcommand{\x}{\xi}              \newcommand{\X}{\Xi}
\renewcommand{\x}{\xi}              \renewcommand{\X}{\Xi}
\newcommand{\y}{\upsilon}       \newcommand{\Y}{\Upsilon}
\newcommand{\z}{\zeta}
%************************************************************************

%LETTRES SCRIPTES
\renewcommand{\AA}{{\cal A}}
\newcommand{\BB}{{\cal B}}
\newcommand{\CC}{{\cal C}}
\newcommand{\DD}{{\cal D}}
\newcommand{\EE}{{\cal E}}
\newcommand{\FF}{{\cal F}}
\newcommand{\GG}{{\cal G}}
\newcommand{\HH}{{\cal H}}
\newcommand{\II}{{\cal I}}
\newcommand{\JJ}{{\cal J}}
\newcommand{\KK}{{\cal K}}
\newcommand{\LL}{{\cal L}}
\newcommand{\MM}{{\cal M}}
\newcommand{\NN}{{\cal N}}
\newcommand{\OO}{{\cal O}}
\newcommand{\PP}{{\cal P}}
\newcommand{\QQ}{{\cal Q}}
\renewcommand{\SS}{{\cal S}}
\newcommand{\RR}{{\cal R}}
\newcommand{\TT}{{\cal T}}
\newcommand{\UU}{{\cal U}}
\newcommand{\VV}{{\cal V}}
\newcommand{\WW}{{\cal W}}
\newcommand{\XX}{{\cal X}}
\newcommand{\YY}{{\cal Y}}
\newcommand{\ZZ}{{\cal Z}}
%***********************************************************************

%SIGNES SPECIAUX
\newcommand{\wh}[1]{\widehat{#1}}
\newcommand{\fd}[1]{\frac{\d}{\d #1}}
\newcommand{\ch}{\widehat{C}}
\newcommand{\gh}{\widehat{\gamma}}
\newcommand{\W}{W_{i}}
\newcommand{\na}{\nabla}
\newcommand{\xint}{\dint d^4x\;}
\newcommand{\sla}{\raise.15ex\hbox{$/$}\kern -.57em}
\newcommand{\Sla}{\raise.15ex\hbox{$/$}\kern -.70em}
\def\h{\hbar}
\def\Lp{\displaystyle{\biggl(}}
\def\Rp{\displaystyle{\biggr)}}
\def\LP{\displaystyle{\Biggl(}}
\def\RP{\displaystyle{\Biggr)}}
\newcommand{\lp}{\left(}\newcommand{\rp}{\right)}
\newcommand{\lc}{\left[}\newcommand{\rc}{\right]}
\newcommand{\lac}{\left\{}\newcommand{\rac}{\right\}}
\newcommand{\identity}{\bf 1\hspace{-0.4em}1}
\newcommand{\complex}{{\kern .1em {\raise .47ex
\hbox {$\scriptscriptstyle |$}}
      \kern -.4em {\rm C}}}
\newcommand{\real}{{{\rm I} \kern -.19em {\rm R}}}
\newcommand{\rational}{{\kern .1em {\raise .47ex
\hbox{$\scripscriptstyle |$}}
      \kern -.35em {\rm Q}}}
\renewcommand{\natural}{{\vrule height 1.6ex width
.05em depth 0ex \kern -.35em {\rm N}}}
\newcommand{\tint}{\int d^4 \! x \, }
\newcommand{\intg}{\int d^D \! x \, }
\newcommand{\intm}{\int_\MM}
\newcommand{\tr}{{\rm {Tr} \,}}
\newcommand{\half}{\dfrac{1}{2}}
\newcommand{\f}{\frac}
\newcommand{\pa}{\partial}
\newcommand{\pad}[2]{{\frac{\partial #1}{\partial #2}}}
\newcommand{\fud}[2]{{\frac{\delta #1}{\delta #2}}}
\newcommand{\dpad}[2]{{\displaystyle{\frac{\partial #1}{\partial
#2}}}}
\newcommand{\dfud}[2]{{\displaystyle{\frac{\delta #1}{\delta #2}}}}
\newcommand{\dfrac}[2]{{\displaystyle{\frac{#1}{#2}}}}
\newcommand{\dsum}[2]{\displaystyle{\sum_{#1}^{#2}}}
\newcommand{\dint}{\displaystyle{\int}}
\newcommand{\eg}{{\em e.g.,\ }}
\newcommand{\Eg}{{\em E.g.,\ }}
\newcommand{\ie}{{{\em i.e.},\ }}
\newcommand{\Ie}{{\em I.e.,\ }}
\newcommand{\nb}{\noindent{\bf N.B.}\ }
\newcommand{\etal}{{\em et al.}}
\newcommand{\etc}{{\em etc.\ }}
\newcommand{\via}{{\em via\ }}
\newcommand{\cf}{{\em cf.\ }}
\newcommand{\twiddle}{\lower.9ex\rlap{$\kern -.1em\scriptstyle\sim$}}
\newcommand{\qed}{\vrule height 1.2ex width 0.5em}
\newcommand{\grad}{\nabla}
\newcommand{\bra}[1]{\left\langle {#1}\right|}
\newcommand{\ket}[1]{\left| {#1}\right\rangle}
\newcommand{\vev}[1]{\left\langle {#1}\right\rangle}
\newcommand{\wt}[1]{\widetilde{#1}}
%***************************************************************************

%EQUATIONS
\newcommand{\equ}[1]{(\ref{#1})}
\newcommand{\be}{\begin{equation}}
\newcommand{\ee}{\end{equation}}
\newcommand{\eqn}[1]{\label{#1}\end{equation}}
\newcommand{\eea}{\end{eqnarray}}
\newcommand{\bea}{\begin{eqnarray}}
\newcommand{\eqan}[1]{\label{#1}\end{eqnarray}}
\newcommand{\ba}{\begin{array}}
\newcommand{\ea}{\end{array}}
\newcommand{\eqac}{\begin{equation}\lac\begin{array}{rcl}}
\newcommand{\eqacn}[1]{\end{array}\right.\label{#1}\end{equation}}
\newcommand{\qq}{&\qquad &}
\renewcommand{\=}{&=&} %seems not to work in footnotes!!!
%---------------  FIN  --------------%
%********************************************************
%%\setlength{\baselineskip}{6ex}
%\setlength{\textwidth}{18cm}
%\setlength{\textheight}{26cm}
%%\setlength{\oddsidemargin}{-1cm}
%%\setlength{\evensidemargin}{-1cm}
%\setlength{\topmargin}{-2cm}
%{\large %POUR PREPRINT
%*********************************************************************
\newcommand{\cb}{{\bar c}}
\newcommand{\mn}{{\m\n}}
\newcommand{\pic}{$\spadesuit\spadesuit$}
\newcommand{\?}{{\bf ???}}
\newcommand{\Tr }{\mbox{Tr}\ }
\newcommand{\adot}{{\dot\alpha}}
\newcommand{\bdot}{{\dot\beta}}
\newcommand{\gdot}{{\dot\gamma}}

\newcommand{\df}[1]{\lp\pa_3\vf\rp_{#1}}
\newcommand{\dft}[1]{\lp \wt{\pa_3\vf}\rp_{#1}}

%%%%%%%%% begin document %%%%%%%%%%%%%%%%%%%%%%%%%%%%%%%%%%%%%%%%
\global\parskip=4pt
\titlepage  \noindent
{
   \noindent
%\today

\hfill GEF-TH-01/2010

%\hfill LPTHE/??-?? 

\vspace{2cm}

\noindent
{\bf
{\large Maxwell--Chern--Simons Theory With Boundary
}}

\vspace{.5cm}
\hrule

\vspace{2cm}

\noindent
\begin{center}{\bf 
A. Blasi, N. Maggiore, N. Magnoli }
\footnote{alberto.blasi@ge.infn.ge, nicola.maggiore@ge.infn.it,
nicodemo.magnoli@ge.infn.it}\end{center}

\noindent
\begin{center}{\footnotesize {\it
 Dipartimento di Fisica \\ Universit\`a di Genova --
via Dodecaneso 33 \\ I-16146 Genova -- Italy \\and \\INFN, Sezione di
Genova 
} }
\end{center}

\begin{center}{\bf S. Storace }
\footnote{stefano.storace@nyu.edu}\end{center}

\begin{center}{\footnotesize {\it
Department of Physics\\
New York University\\
4 Washington Place\\
New York, NY 10003, USA
} }
\end{center}

\vspace{1cm}
\noindent
{\tt Abstract~:}
The Maxwell--Chern--Simons (MCS) theory with planar boundary is considered. 
The boundary is introduced according to Symanzik's basic principles 
of locality and separability. A method of investigation is proposed, 
which, avoiding the straight computation of correlators, is appealing 
for situations where the computation of propagators, modified 
by the boundary, becomes quite complex. For MCS theory, the outcome is that 
a unique solution exists, in the form of chiral conserved currents, 
satisfying a Ka\v{c}--Moody algebra, whose central charge does not depend 
on the Maxwell term.

\vfill\noindent
{\footnotesize {\tt Keywords:}
Quantum Field Theory ,
Boundary Conditions, Boundary Algebra
% BRST Quantization,
% Chern-Simons Theories,
% Topological Field Theories,
% Noncommutative Field Theories.
\\
{\tt PACS Nos:} 
03.70.+k Theory of Quantized Fields,
11.15.Wx	Topologically massive gauge theories
11.15.Yc Chern--Simons gauge theory
%11.40.Ex	Formal properties of current algebras
% 11.15.-q Gauge Field Theories,
% 11.10.Gh Renormalization,
% 11.10.Nx Noncommutative Field Theory.
}
\newpage
%*****************************************************
%%%%%%%%%% end of title page %%%%%%%%%%%%%%%%%%%%%%%%%%%%%%
\begin{small}
%\tableofcontents
\end{small}

\setcounter{footnote}{0}

%\tableofcontents

\section{Introduction}

In the last decades, there has been vast interest in the
role of the space boundary in Quantum Field Theory (QFT)
 \cite{Symanzik:1981wd,Witten:1988hf,Moore:1989yh}. 
The main reason is that
the presence of a 
boundary changes the properties of a physical
system, originating phenomena like the Casimir 
Effect  \cite{Casimir:1948dh} or the edge states in the 
Fractional Quantum Hall Effect 
 \cite{Wenbook}.

In some cases, the boundary even represents the center
of the investigation, as in the case of black holes  \cite{Banados:1992wn}, where
the horizon of events can be indeed regarded as a boundary.
Boundary effects play an important role also 
in critical phenomena \cite{Diehl}.
%behaviour of magnets and alloys with free surfaces  \cite{Dosch}
%and critical absorption of fluids on walls and interfaces  \cite{Liu}.
\newline

In particular, the effect of a boundary 
is extremely interesting in
topological field theories.
Indeed, topological theories are known to have no
local observables except when the base
manifold has a boundary  \cite{Witten:1988hf,Moore:1989yh}.
Our attention will be actually focused
on the topological, three dimensional 
Chern--Simons (CS) model 
 \cite{Witten:1988ze,Birmingham:1991ty}. 

The introduction of a boundary
in CS theory has been
investigated in previous works  \cite{Witten:1988hf,Moore:1989yh},
with the result that the local observables arising from the
presence of the boundary
are two-dimensional conserved chiral 
currents generating the Ka\v{c}-Moody algebra 
 \cite{Kac:1967jr}
of the Wess-Zumino-Witten model  \cite{Witten:1983ar}.
In other words,the boundary establishes a strong connection 
between CS theory and conformal 
field theories  \cite{Belavin:1984vu}.

An analogous result has been obtained for another
topological field theory, namely the
topological BF model
 \cite{Deser}
in three dimensions,
for which it has been shown 
as well the existence of chiral currents
living on the boundary
and satisfying a Ka\v{c}-Moody 
algebra  with central extension  \cite{Maggiore:1992ki}.
\newline

The existence
of a Ka\v{c}-Moody algebra
with central extension in the CS model with boundary
has led to interesting applications due to the connection
of the CS theory with several physical systems.

In fact, following the equivalence between CS theory
and (2+1) dimensional gravity theory with a cosmological constant
 \cite{Witten:1988hc,Deser:1983dr}, it has been stressed that the algebra
 plays a crucial role in understanding
the statistical origin of the entropy of a black hole  \cite{Banados:1992wn}.
It is thus possible to use the algebra to compute  
the BTZ black hole (negative cosmological constant)
entropy \cite{Banados:1998ta}
and the Kerr-de Sitter space (positive cosmological constant)
entropy  \cite{Park:1998qk}.

In condensed matter physics,
the abelian CS model provides 
an effective low energy theory for the Fractional
Quantum Hall Effect  \cite{Wenbook}.
Further, when a boundary is taken into account,
 the Ka\v{c}-Moody algebra 
describes the boundary
chiral currents which are indeed observed 
on the edge of the Hall bar 
(edge states) \cite{Wenbook}.
\newline

On the other hand, the coupling of the CS model
to other theories also gives rise to interesting results. 
For instance, it can be coupled to fermion or boson fields to attach
magnetic flux to charge density, thus providing
 an explicit realization of \emph{anyons}, \emph{i.e.} particles
living only in systems with two spatial dimensions and
satisfying a fractional statistics  \cite{Wilczek:1990ik}. 

Among the others, one of the most striking properties
of the CS term is that, when added to the 
three dimensional Yang-Mills
action, it originates a 
\emph{topologically massive gauge theory}  \cite{Deser:1981wh}. 

Its abelian version, 
the Maxwell--Chern--Simons (MCS) theory, when 
coupled with fermions
defines
a three dimensional 
modified electrodynamics, in which the ``photons''
are massive and have a single state of helicity  \cite{Deser:1981wh}. 
The Casimir effect for topologically massive electrodynamics
could provide, in principle, a way to ``measure'' the topological mass 
of the ``photons''  \cite{Milton:1990yj}.

Moreover, the addition of a Maxwell term to the effective 
low energy theory for the Quantum Hall Effect 
allows the description of the gap
between the ground state and the bulk 
elementary excitations  \cite{Wenbook}.

The Maxwell theory
coupled to the CS action on a manifold with boundary has 
a further application in (2+1) dimensional
quantum gravity: the so called
Einstein--Maxwell--Chern--Simons theory  \cite{Andrade:2005ur}. 
In this framework, the boundary
can be regarded as the horizon of the 
black hole solutions, and 
the gauge field coupled to gravity
describes a topologically 
massive electromagnetic field which
provides the black hole with an electric charge,
which, coming from the CS term, is 
``kind of'' \emph{topological}. The result
is thus called ``charged black hole''.
\newline

On the other hand, the Maxwell model is not topological, and therefore
the question if its addition spoils the chiral current algebra of the CS theory
arises naturally: the aim of this paper is actually to discuss this issue. 

To reach this task, we first must face a further
problem, \emph{i.e.} the method. Indeed, the inclusion of a boundary 
in field theory is a highly non trivial
task if one wishes to preserve locality and power counting, the most
basic ingredients of QFT. 

In 1981
K. Symanzik  \cite{Symanzik:1981wd} addressed this question: 
his key idea was to add to the
bulk action a local boundary term which modifies the propagators of the fields
in such a way that nothing propagates from one side of the boundary
to the other. He called this property ``separability'' and showed that
it requires the realization of a well identified 
class of boundary conditions that can be implemented by a local bilinear
interaction.
\newline

These ideas strongly inspired the authors of  \cite{Blasi:1990bk}, 
\cite{Emery:1991tf} and \cite{BFMMS}, who
used a closely related approach to compute the chiral current algebra 
living on the boundary of the three-dimensional
CS model. 

Indeed, the authors of  \cite{Blasi:1990bk} added to the action
local boundary terms compatible with power counting, using a covariant
gauge fixing. 
In  \cite{Emery:1991tf,BFMMS}, on the other hand, a regularization-free procedure
was followed: the equations
of motion, rather than the action, were modified by appropriate boundary terms,
and a non-covariant axial gauge was preferred. 

The main reason for the latter choice was that
the main advantage of a covariant gauge, \emph{i.e.} Lorentz invariance, already fails
due to the presence of the boundary. On the other hand, the axial choice does not 
completely fix the gauge  \cite{Soldati:1991book}, and a residual gauge invariance exists,
implying the existence 
of a Ward identity which plays a crucial role
since, when restricted to the boundary, it might
generate a chiral current algebra, as we shall see. 
\newline

In both these works, the explicit computation of the propagators of the
 theory seems to be necessary. However, this step could be quite difficult
in other theories, like MCS, and 
another way of investigation is worthy, which avoids
the explicit computation of the Green functions of the
theory. 

This is precisely the choice that we make. Our approach is actually more similar
to that used in  \cite{Emery:1991tf,BFMMS}, since we focus on the effect of the boundary on the equations
of motion rather than on the action, and we adopt
a noncovariant axial gauge rather than a covariant one.
\newline

The basic idea is that, after modifying the equations of motion
by means of
 boundary terms satisfying general basic requirements, we 
integrate them 
in proximity of the boundary and use Symanzik's idea of separability
to $determine$, rather than $impose$, the boundary conditions on the propagators, that can be
expressed as boundary conditions on the fields  \cite{Symanzik:1981wd}. 
These, in turn, have
an effect on the boundary breaking term of 
the residual Ward identity that generates the
Ka\v{c}-Moody algebra living on the boundary.
\newline

For these reasons, our method is suitable also for
a different kind of investigation.  
Indeed, the approach in 
 \cite{Park:1998yw} and  \cite{Dunne:1990hh} 
is good to describe 
just a portion of space: there 
is not a \emph{beyond the boundary}.
This is not a 
problem for the description of systems like 
electrodynamics on a disk  \cite{Balachandran:1993tm} or
a Hall bar in the framework of the 
Quantum Hall Effect  \cite{Wenbook}, 
in which one is interested in the dynamics of the internal
system only.

However, the description of what there is beyond the boundary
is fundamental in the physics of {\it defects}  \cite{Mermin:1979zz}, and our
approach is actually suitable for this task.
In the study of semiconductors  \cite{defectbook},
for instance, it is important to be able to describe
local interactions between the particles 
and the imperfections of
the material, approximated by $\delta$-type interactions.

On the other hand, topological defects are gathering
more and more importance also in the framework of
astroparticle physics, where they are proposed
as an explanation for the formation
of cosmic structures
and for the generation of extremely high-energy
cosmic rays  \cite{Berezinsky:1998ft,Aharonian:1992qf}. 

Even 
considering an impenetrable boundary, we keep
 both the left and the right 
side of spacetime. Indeed, we
can fix the parameters
of our general description in such a way to
decouple the opposite sides of spacetime completely, 
and therefore in principle we are able to
examine just one of the two sides, if we wish. 
In a certain way, this is actually 
what we will do by imposing the conservation of
the bulk discrete symmetry \equ{eqn:inversion} 
which we called parity. 

In other words, according to Symanzik's approach, the boundary 
is {\it defined} by the decoupling condition, which prevents 
correlations between points belonging to opposite sides of the 
boundary. A defect could be treated by using the same approach 
described in this paper, by simply relaxing the decoupling 
condition. In this way, the resulting modified theory describes 
interactions between a bulk and a $\delta$-type insertion in the 
action, {\it i.e.} a defect, as it is done in 
 \cite{Mintchev:2007qt}.

We organized this paper as follows. 
In Section 2 we reconsider the abelian
CS theory with boundary, and we show that 
our method leads again to the correct 
results found in  \cite{Blasi:1990bk}
and  \cite{Emery:1991tf}
for both the boundary conditions and the residual
Ward identity, and we illustrate how the Ward identity
generates the Ka\v{c}-Moody algebra. 

Although the method illustrated in Section 2
cannot be entirely found elsewhere already, 
the main original results can be found in Section 3, 
where we introduce the Maxwell
term in the CS theory. We follow the 
same scheme of Section 2 to investigate 
whether, also in this case, conserved chiral currents exist
which satisfy some kind of algebra, like in the pure CS theory.
We stress again that the answer to this question is not at all to be
taken for granted, due to the
non-topological character of the bulk theory.

Our concluding remarks summarize our results
and draw some conclusions, with some further
suggestions of possible applications and extensions.

\section{``Boundarization'': the pure Chern--Simons case}
\subsection{The model}

In this Section we recall 
some known results concerning CS theory with planar boundary 
\cite{Blasi:1990bk,Emery:1991tf,BFMMS}. This will give us the 
opportunity to illustrate the method we shall follow in what really 
matters us, that is the MCS theory with boundary, 
in order to investigate the boundary physics without calculating 
explicitly the correlators of the theory in presence of the boundary.

In euclidean space, the abelian CS action reads~:
\be
S_{cs}= -\frac{k}{2}\int d^{3}x \epsilon_{\m\n\r}
A_{\m}\pa_{\n}A_{\r} \ .
\eqn{eqn:CSaction}
It will be convenient to
work in euclidean light-cone coordinates,
defined as: 
\be \ba{rcl}
u &=& x_{2}\\
z &=& \frac{1}{\sqrt{2}}({x_1-ix_0})\\
\bar{z} &=&\frac{1}{\sqrt{2}}({x_1 +ix_0})\ ,
\ea 
\eqn{eqn:lc}
which induce similar definitions in the space of the fields:
\be\ba{rcl}
A_{u}&=& A_{2}\\
A&=& \frac{1}{\sqrt{2}}(A_{1}+iA_{0})\\
\bar{A}&=& \frac{1}{\sqrt{2}}(A_{1}-iA_{0})
\ea  \ .
\eqn{eqn:lcfields}
The CS action then reads
\be
S_{cs}=-k\int dudzd\bar{z}\
\left(
\bar{A}\partial_{u}A + 
A_{u}\partial\bar{A} - A_{u}\bar\partial A
\right)\ .\label{lccs}
\ee
The gauge fixing term is~:
\be
S_{gf}=-\int dudzd\bar{z}\ A_u b \ ,\label{lcgf}
\ee
which corresponds to the axial gauge choice
\be
A_{u}=0\ .\label{axialgauge}
\ee
The complete action is
\be
\S_{CS}=S_{cs}+S_{gf}+S_{ext} \ ,
\ee
where in $S_{ext}$ we coupled, as usual, external sources 
$J_{\Phi}$ to the quantum fields $\Phi=A_{u},A,\bar{A},b$~:
\be
S_{ext}= -\int du dz d\bar{z}\
\sum_{\Phi}J_{\Phi}\Phi\ .\label{lcext}
\ee

Besides the canonical mass dimensions, the fields appearing in the 
action $\S_{CS}$ are assigned an additional quantum number, called
$helicity$, which encodes the two--dimensional Lorentz invariance of 
the theory on the planes $u=constant$.

The dimensions and helicities of the fields and sources are 
summarized in Table~1~:
$$
\stackrel{Table\;1\;:\;CS\;Quantum\;Numbers}{
\begin{tabular}{|c|c|c|c|c|c|c|c|c|c|c|c|c|c|c|}
\hline 
& $ $ & $ $ & $ $ & $ $ & $ $ & $ $ & $ $ & $ $ & $ $ & $ $ & $ $ & $ $ & $ $ & $ $ \\
& $A_{u}$&$A$ & $\bar{A} $ & $b$ & $J_{u}$&$J$ & $\bar{J} $ & $J_{b}$ & $\partial_{u}$ & $\partial$ &
$\bar{\partial}$ & $u$ &$z$ &$\bar{z}$\\ \hline
$\dim $ & $1$&$1$ & $1$ & $2$ & $2$&$2$ & $2$ & $1$ & $1$ &$1$& $1$ &$-1$&$-1$&$-1$ \\
\hline
$\mathrm{hel}$ & $0$&$1$ & $-1$ & $0$ & $0$&$1$ & $-1$ & $0$ & $0$&$1$ & $-1$ & $0$ &$-1$ & $+1$\\ \hline 
\end{tabular}
}  $$

Moreover, as it is well known, the axial gauge choice 
\equ{axialgauge} does not completely fix the gauge: a residual
local gauge invariance remains, which is expressed by the local Ward 
identity 
\be
\partial \bar{J}+\bar{\partial}J +\partial_{u}
J_{u}+\partial_{u}b=0
\label{eqn:CSlocward} \ ,
\ee
which, once integrated, gives
\be \int du \lp\pa \bar{J}+\bar{\pa} J\rp=0 \ .\eqn{eqn:WardZZ}
Finally, the set of symmetries of the action $\S_{CS}$ is completed by
the discrete transformation involving at
the same time coordinates and fields:
\be  \ba{rcl}
	z &\leftrightarrow& \bar{z} \\
	u &\rightarrow& -u  \\
	A &\leftrightarrow& \bar{A} \\
	A_{u} &\rightarrow& - A_{u} \\
	b &\rightarrow& -b \\
        J &\leftrightarrow& \bar{J} \\
        J_u &\rightarrow& -J_u \\
        J_b &\rightarrow& -J_b \ ,
        \ea \eqn{eqn:inversion}
which we will refer to by using the term \emph{``parity''}.

The field equations of motion are
\bea
k\left(\bar{\partial}A_{u}-\partial_{u}\bar{A}\right)+\bar{J}&=&0 \\
k\left(\partial_{u}A-\partial A_{u}\right)+J&=&0  \\
k\left(\partial \bar{A}-\bar{\partial}A+\f{1}{k}b
\right)+J_{u}&=&0 \\
A_{u}+J_{b}&=&0    \label{eqn:CSIV}\ .
\eea

\subsection{The boundary}

We now want to introduce a boundary in the
theory, and we choose the planar surface $u=0$.

The effect of the boundary is to break
the original 
equations of motion, in a way which must respect the following basic 
constraints.

{\bf Locality:} The
 boundary contribution must be local. This means that all possible breaking
terms have the form 
\be \d^{(n)}(u)X(z,\bar{z},u) \ , \ee
where $\d^{(n)}(u)$ is the $n$-order derivative of the 
Dirac delta function with
respect to its argument, and 
$X(z,\bar{z},u)$ is a local functional.

{\bf Separability:} This constraint, called also {\it decoupling 
condition}, refers
to Symanzik's original idea  \cite{Symanzik:1981wd} 
according to which 
the $n$-point Green 
functions which involve two fields
computed in points belonging to opposite sides of space must vanish.
In particular, the propagators
of the theory must satisfy~:
\be uu'<0\Rightarrow \D_{AB}(x-x')=\langle T\lp\vf_{A}(x)\vf_{B}(x')\rp\rangle=0 \ 
, \ee
where
\be
x = (z,\bar{z},u)\ ,
\ee
and $\vf_{A}(x)$ is a generic field of the theory.
This property is satisfied by propagators 
which split according to
\be \D_{AB}(x-x')=\theta_+\D_{AB+}(x-x')+\theta_-\D_{AB-}(x-x') \ , 
\eqn{eqn:scaldec}
where $\D_{AB+}(x-x')$ and $\D_{AB-}(x-x')$ are respectively the propagators for the 
right and the left side of spacetime,
\be \theta_\pm = \theta(\pm u)\theta(\pm u') \ee and
$\theta(x)$ is the step function, defined as usual.

The relation \equ{eqn:scaldec} is referred to as \emph{``decoupling 
condition''}, which induces the decomposition of the generating 
functional $\WW$ of connected Green functions according to
\be \WW=\WW^++\WW^- \ , \eqn{eqn:scalgenfuncbrk}
where $\WW^+$ and $\WW^-$ are the generators of the connected Green functions 
for the right side and the left side of spacetime, respectively.

{\bf Linearity:} Finally, we require that the boundary contributions
to the equations of motion be linear in the fields because, in general,
 the symmetries 
of the classical action, if only linearly broken at the classical level,
nonetheless remain exact symmetries of the quantum action  \cite{Weinberg1}. 
Moreover, we are considering a free field theory and therefore nonlinear terms
in the field must not occur in the equations of motion.

The most general 
broken equations of motion
which satisfy all the above requirements are
\bea
k\left(\bar{\partial}A_{u}-\partial_{u}\bar{A}\right)+\bar{J}&=&\d(u)(c_1^+\bar{A}_++c_1^-\bar{A}_-)\label{1}\\
k\left(\partial_{u}A-\partial 
A_{u}\right)+J&=&\d(u)(c_2^+A_++c_2^-A_-)\label{2} \\
k\left(\partial \bar{A}-\bar{\partial}A+\f{1}{k}b
\right)+J_{u}&=&\d(u)(c_3^+{A_u}_++c_3^-{A_u}_-)\label{3}\\
A_u+J_{b}&=&0\label{4}  \ ,
\eea
where  $c_i^\pm$ are constant parameters, and 
$A_\pm(Z)$, $\bar{A}_\pm(Z)$ and ${A_u}_\pm(Z)$ are the boundary fields on the right, respectively on the left, of the boundary:
\bea
A_\pm(Z)&=&\lim_{u\rightarrow0^\pm}{A(x)} \\
\bar{A}_\pm(Z)&=&\lim_{u\rightarrow0^\pm}{\bar{A}(x)}\\
{A_u}_\pm(Z)&=&\lim_{u\rightarrow0^\pm}{A_u(x)} \ ,
\eea
with $Z = (z,\bar{z})$. Under parity
\equ{eqn:inversion}, the boundary fields transform according to
\bea 
A_{\pm}&\leftrightarrow&\bar{A}_{\mp} \\
{A_u}_\pm&\rightarrow&-{A_u}_\mp \ .
\eea
Imposing the parity constraint \equ{eqn:inversion}
on the broken equations of motion leads to
\bea 
c_1^\pm&=&c_2^\mp \\ 
c_3^+&=&-c_3^- = c_3 \ .
\eea

In addition, the broken equations of motion must satisfy what in 
\cite{Emery:1991tf} has been called $compatibility$, which is nothing 
else than the commutation property which equations of motion in 
general obey. This algebraic constraint further reduces the number of 
parameters appearing in the r.h.s. of the boundary equations of 
motion~:
\be 
c_1^+=c_1^- =  c \ .
\ee

Finally, the equations of motion, broken by the presence of the 
planar boundary $u=0$,  which 
respect parity and compatibility are
\bea
k\left(\bar{\partial}A_{u}-\partial_{u}\bar{A}\right)+\bar{J}&=&
c\d(u)(\bar{A}_++\bar{A}_-) \label{eqn:CSbrokeqI}\\
k\left(\partial_{u}A-\partial A_{u}\right)+J&=&c\d(u)(A_++A_-)\label{eqn:CSbrokeqII}\\
k\left(\partial \bar{A}-\bar{\partial}A+\f{1}{k}b
\right)+J_{u}&=&c_3\d(u)({A_u}_+-{A_u}_-) \label{eqn:CSbrokeqIII}\\
A_u+J_{b}&=&0   \label{eqn:CSbrokeqIV}\ .
\eea

\subsection{The boundary conditions}

In QFT  with boundary, a crucial issue is represented by the 
boundary condition for the quantum fields. This problem is often 
solved by imposing ``by hand'' boundary conditions and see what 
happens. In Symanzik's approach, the only constraint is the 
decoupling condition, which concerns the 
Green functions, as we discussed previously. The boundary conditions 
on the fields, rather than being imposed, are read, once the basic 
separability condition on propagators is realized.

The price to pay is, of course, the need of an explicit computation 
of the propagators, which, in presence of a boundary, is not at all 
an easy task, usually.

For CS theory the problem has been solved in 
\cite{Blasi:1990bk} and 
in \cite{Emery:1991tf}, by two quite different approaches, and, from 
the expression of
the propagators on the boundary, the boundary conditions on the 
fields --which turn out to be of the Dirichlet type -- are 
inferred. But, for more complicated situations, like the one this 
paper is devoted to, this program might be out of reach. 

Here, we present an easy way out. That is, we claim to be able to 
carry out Symanzik's program for QFTs with boundary, and find out the 
boundary physics, without explicitly computing the Green functions.

In this Section we are treating the pure CS case, which 
is known, just to illustrate the idea. 

Definitely more interesting is the adoption of our simple, but 
powerful, method, to the MCS theory, what will 
be done in the next Section.

After setting the sources to zero, let us integrate \equ{eqn:CSbrokeqI} and \equ{eqn:CSbrokeqII}
with respect to  $u$ from $-\e$ to $\e$.
Taking into account also \equ{eqn:CSbrokeqIV}, we get
\bea \lp k-c \rp \bar{A}_-&=& \lp k+c \rp \bar{A}_+\label{eqn:boundcondI}\\
     \lp k+c \rp A_-&=&\lp k-c \rp A_+ \label{eqn:boundcondII}\ .\eea
Due to the decoupling condition, each side of the above identities
must vanish separately:
\bea
\lp k-c \rp \bar{A}_-&=&0\\
\lp k+c \rp \bar{A}_+&=&0\\
\lp k+c \rp A_-&=&0\\
\lp k-c \rp A_+&=&0 \ .
\eea
The nontrivial solutions are given by 
\bea c=k\Rightarrow \bar{A}_+=&0&=A_- \label{eqn:CSchoiceI}\\
     c=-k\Rightarrow \bar{A}_-=&0&=A_+ \ .\label{eqn:CSchoiceII} \eea
This is exactly the same result previously found in 
 \cite{Emery:1991tf,BFMMS,Blasi:1992mm}: the fields obey
\emph{Dirichlet} boundary conditions 
on both sides of the dividing plane $u=0$.

The curious reader is invited to compare how the same result has ben 
obtained in \cite{Blasi:1990bk} and 
\cite{Emery:1991tf}, where, like we did, the boundary conditions are not put 
``by hand'' in the theory but derived from the Symanzik's decoupling 
condition.

It is clear that the choices 
\equ{eqn:CSchoiceI} and \equ{eqn:CSchoiceII} essentially describe the same physics, 
since they are related by the parity transformation \equ{eqn:inversion}.
Therefore, in what follows we choose the solution with $c=k$, keeping
in mind that a kind of ``mirror solution'' exists. 

The equations of motion thus become
\bea
k\left(\bar{\partial}A_{u}-\partial_{u}\bar{A}\right)+\bar{J}&=&k\d(u)\bar{A}_- \\
k\left(\partial_{u}A-\partial A_{u}\right)+J&=&k\d(u)A_+\\
k\left(\partial \bar{A}-\bar{\partial}A+\f{1}{k}b
\right)+J_{u}&=&c_3\d(u)({A_u}_+-{A_u}_-) \\
A_u+J_{b}&=&0 \ .
\eea
Correspondingly, the local Ward identity \equ{eqn:CSlocward}  acquires a
boundary breaking:
\be
\pa \bar{J}+\bar{\pa}J +\pa_u
J_{u}+\pa_u b=k \d(u)\lp\bar{\pa}A_++\pa
\bar{A}_-\rp + \pa_u\lc c_3\d(u)({A_u}_+-{A_u}_-)\rc\ ,
\eqn{eqn:CSwardbound} 
which yields the integrated Ward identity
\be
\f{1}{k}\int du\lp\pa
\bar{J}+\bar{\pa}
J\rp=\bar{\pa}A_++\pa\bar{A}_- \ .
\eqn{eqn:CSwardboundint}

\subsection{The boundary algebra}

In this Section we recall how the Ward identity \equ{eqn:CSwardboundint}
implies the existence of a Ka\v{c}-Moody algebra 
of conserved chiral currents
on the boundary.

We can rewrite the Ward identity in a functional way as
\be
\f{1}{k}\int du\lc\pa
\bar{J}(x)+\bar{\pa}
J(x)\rc=\bar{\pa}\left.\f{\d \WW_+}{\d \bar{J}(x)}\right|_{u=0^+}+
\pa\left.\f{\d \WW_-}{\d J(x) }\right|_{u=0^-} \ .
\ee
We then differentiate with respect to
$\bar{J}(x')$, with $x'$ lying on the right side of space next to the boundary,
 and then set the sources to zero. We thus get
\bea
\f{1}{k}\int du
\pa\d^3(x-x')&=&\left.\lp\bar{\pa}\left.\f{\d^2 \WW_+}{\d \bar{J}(x') \d \bar{J}(x)}\right|_{u',u=0^+}\rp \right|_{J_\Phi=0}
\nonumber\\
&+&\left.\lp\pa\left.\f{\d^2 \WW_-}{\d \bar{J}(x') \d J(x)}\right|_{u'=0^+,u=0^-}\rp\right|_{J_\Phi=0} \ . 
\eea
Recalling that $\WW_\pm$ are the generators of the connected Green functions
for the $u>0$ and $u<0$ sides of spacetime respectively, 
the right-hand side of this expression involves
propagators, the second of which vanishes due to the decoupling condition. 
Therefore, we are left with
\be
 \f{1}{k}\pa\d^2(Z-Z')=\bar{\pa}\vev{T(A_+(Z')A_+(Z))} \ .
\eqn{eqn:CSwardprop}
Keeping in mind the definition of the time-ordering operator $T$, we have
to specify the role of time in light-cone variables. This issue             
has already been extensively studied 
in the literature  \cite{Dirac:1949cp}.
The outcome is that a possibility 
is to identify the light-cone variable $\bar{z}$ with time.                               

We can now explicitly compute the
right-hand side of \equ{eqn:CSwardprop}:
\bea
&&\bar{\pa}\vev{T(A_+(Z')A_+(Z))}=\nonumber\\
&&\nonumber\\
&=&\bar{\pa}\vev{\theta(\bar{z}'-\bar{z})A_+(Z')A_+(Z)+
\theta(\bar{z}-\bar{z}')A_+(Z)A_+(Z')}\nonumber\\
&=&-\vev{\d(\bar{z}'-\bar{z})A_+(Z')A_+(Z)+\theta(\bar{z}'-\bar{z})A_+(Z')\bar{\pa}A_+(Z)}\nonumber\\
&&+\vev{\d(\bar{z}-\bar{z}')A_+(Z)A_+(Z')+\theta(\bar{z}-\bar{z}')\bar{\pa}A_+(Z)A_+(Z')} \nonumber\\
&&\nonumber\\
&=&\vev{\theta(\bar{z}'-\bar{z})A_+(Z')\bar{\pa}A_+(Z)+
\theta(\bar{z}-\bar{z}')\bar{\pa}A_+(Z)A_+(Z')}\nonumber\\
&&+\d(\bar{z}-\bar{z}')\vev{\lc A_+(Z),A_+(Z')\rc} \ . \label{eqn:CSboundcorr}
\eea
On the other hand, after setting the sources to zero, the Ward identity \equ{eqn:CSwardboundint}
and the decoupling condition yield the
\emph{chirality condition}  \cite{Blasi:1990bk,Emery:1991tf,BFMMS}:       
\bea
0&=&\bar{\pa}A_+ \Rightarrow A_+=A_+\lp z\rp\label{eqn:CSchirality}\\
0&=&\pa\bar{A}_- \Rightarrow \bar{A}_-=\bar{A}_-\lp \bar{z}\rp
\eea
Substituting \equ{eqn:CSchirality} into \equ{eqn:CSboundcorr} and then
into \equ{eqn:CSwardprop} we get
\be
 \f{1}{k}\pa\d^2(Z-Z')=\f{1}{k}\d(\bar{z}-\bar{z}')\vev{\pa\d(z-z')}=
\d(\bar{z}-\bar{z}')\vev{\lc A_+(z),A_+(z')\rc} \ ,
\ee
which finally yields the commutation relation
\be
\lc A_+(z),A_+(z')\rc=\f{1}{k}\pa\d(z-z') \ . \label{eqn:CScommutation}
\ee
This is the abelian counterpart of the
 Ka\v{c}-Moody algebra  \cite{Kac:1967jr} of the Wess-Zumino-Witten model 
 \cite{Witten:1983ar} generated by the 
chiral currents found in  \cite{Moore:1989yh,Blasi:1990bk,Emery:1991tf}. Under 
this respect, the coefficient
$\displaystyle{\f{1}{k}}$ can be seen as the central
charge 
of the Ka\v{c}-Moody algebra. Note that the parity symmetry \equ{eqn:inversion}
implies the mirror algebra
\be
\lc \bar{A}_-(\bar{z}),\bar{A}_-(\bar{z}')\rc=\f{1}{k}\bar{\pa}\d(\bar{z}-\bar{z}') 
\ee
on the opposite side of the boundary.

Moreover, from \equ{eqn:CSchirality} and \equ{eqn:CSchoiceI}, we get
\be
\bar{\pa}A_++\pa\bar{A}_+=0 \ ,\label{eqn:CScontinuity}
\ee
which is the conservation relation
for the planar boundary field, written in light-cone coordinates
 \cite{BFMMS}. Indeed, recalling \equ{eqn:lcfields},
we can come back to the euclidean components of the gauge field:
\bea
A_+& = &\f{1}{\sqrt{2}}\lp A_{1+}+iA_{0+}\rp\nonumber\\
\bar{A}_+& = &\f{1}{\sqrt{2}}\lp A_{1+}-iA_{0+}\rp \ .
\eea
The relation \equ{eqn:CScontinuity} then takes the form
\be
\pa_1A_{1+}+\pa_0A_{0+}=0 \ ,
\eqn{flag}
which is easily identified with a continuity relation involving
a density $A_{0+}$ and a current $A_{1+}$.

Furthermore, from \equ{eqn:CSchoiceI} and \equ{eqn:CSchirality} follows
\be
\pa_1A_{0+}-i\pa_0A_{0+}=0 \ ,
\ee
which, compared to \equ{flag}, implies the identification of $A_{0+}$ and $A_{1+}$
\be
A_{1+}=iA_{0+} \ .
\ee
Consequently, the algebra \equ{eqn:CScommutation} 
can be written in terms of the density $A_{0+}$:
\be
\lc A_{0+}(z),A_{0+}(z')\rc=-\f{1}{2k}\pa\d(z-z') \ .
\ee

Summarizing, in this Section we recovered the general 
result  \cite{Witten:1988hf,Moore:1989yh}
that a topological    
field theory acquires local observables only when
a boundary is introduced. In the CS case, the observables
are conserved chiral currents living on the boundary
and satisfying a Ka\v{c}-Moody algebra, whose central charge
is the inverse of the CS coupling constant, as discussed in
 \cite{Blasi:1990bk,Emery:1991tf,BFMMS}.

Our approach led us to the same results previously obtained in 
\cite{Blasi:1990bk,Emery:1991tf,BFMMS}, but, we stress once again,
with the remarkable difference that we have used an algebraic method which
has allowed to avoid the explicit computation of the propagators of the theory.
This will be extremely useful in cases, like that described in the 
next Section, 
in which the explicit computation of correlators is not that easy.

\section{One step beyond: adding a Maxwell term}

\subsection{The model}

In this Section we study the effect of a planar boundary on the MCS theory. 
We stress that adding a Maxwell term to the CS theory is not a mere extension, 
in particular for what concerns the boundary physics. The Maxwell term, indeed, 
breaks the topological character of the theory, and, hence, all properties 
which are peculiar to topological QFTs in principle do not hold anymore. 
The Maxwell term introduces local degrees of freedom, which are absent 
in topological CS theory, and what happens when a boundary is introduced 
is far from being obvious. 

From the physical point of view, the MCS theory 
describes a massive spin 1 particle with a single
state of helicity, and which  
can be coupled with fermions to define a modified electrodynamics of fermions
interacting with each other and with 
topologically massive ``photons''  \cite{Deser:1981wh}. The introduction of local degrees of freedom renders the MCS theory particularly relevant both in solid state physics (Quantum Hall Effect \cite{Wenbook}) and (black holes \cite{Banados:1992wn}), as explained in the Introduction.

For this reason, it is interesting to investigate wether 
\begin{enumerate}
\item chiral
\item conserved currents exist, which
\item satisfy an algebra, and of which type, and, 
\item if the Ka\v{c}--Moody structure is preserved, how sensitive is the central charge to the Maxwell term.
\end{enumerate}

The MCS action, in euclidean space, is
\be
S_{MCS}=S_{CS}+S_{M}\ ,
\ee
where $S_{CS}$ is the CS action \equ{eqn:CSaction}, $S_{M}$ is 
the Maxwell action
\be
+ \f{\a}{4}\int d^3x F_{\m \n}F_{\m \n}\ ,
\ee
and $F_{\mu\nu}(x)$ is the field strength.

Notice that the theory actually depends only
on one parameter only, that we identify with $k$, which represents
 the topological mass.
In fact, the parameter $\a$ in front of the Maxwell term can be reabsorbed
by a redefinition of the gauge field $A_\m(x)$. Nevertheless, we prefer
to keep it in order to be able, at a later step, to switch off
the Maxwell term and make contact with the pure CS case. 
Another good reason to keep
$\a$, is that in the nonabelian extension it is not possible to reabsorbe
it, and therefore in the general case it is a real coupling constant.

The Maxwell action in light-cone coordinates \equ{eqn:lcfields}, is~:
\bea
S_{M}&=& -\f{i\a}{2}\int du dz d\bar{z} \lc \bar{\pa}A\bar{\pa}A
+\pa\bar{A}\pa\bar{A}-2\pa\bar{A}\bar{\pa}A\right.\nonumber\\
&&\left.+2\pa_uA\bar{\pa}A_u
+2\pa_u\bar{A}\pa A_u-2\pa_u\bar{A}\pa_u A-2\pa A_u\bar{\pa}A_u\rc \label{eqn:lcMaction} \ .
\eea
Recalling that the CS theory in euclidean light-cone coordinates is given by
$S_{CS}$ \equ{lccs}}, that the gauge fixing term for the axial gauge 
$A_u=0$ is $S_{gf}$
\equ{lcgf}, and that the external sources are coupled to quantum 
fields through $S_{ext}$ \equ{lcext}, 
the complete action for the theory that we now consider is
\be
\S_{MCS}=S_{CS}+S_{M}+S_{gf}+S_{ext} \ .\label{mcsaction}
\ee

The presence of the Maxwell term modifies the canonical mass 
dimensions of fields and sources. Table~2 summarizes dimensions and 
helicities of MCS theory. Notice also that the CS coupling constant 
$k$ acquires a mass dimension $[k]=1$.

\bea
&\stackrel{Table\;2\;:\;MCS\;Quantum\;Numbers}{
\begin{tabular}{|c|c|c|c|c|c|c|c|c|}
\hline 
& $ $ & $ $ & $ $ & $ $ & $ $ & $ $ & $ $ & $ $ \\
& $A_{u}$&$A$ & $\bar{A} $ & $b$ & $J_{u}$&$J$ & $\bar{J} $ & $J_{b}$ \\ \hline
$\dim $ & \small{$1/2$}&\small{$1/2$}
 & \small{$1/2$} & \small{$5/2$} & \small{$5/2$}
&\small{$5/2$} & \small{$5/2$} & \small{$1/2$} \\
\hline
$\mathrm{hel}$ & $0$&$1$ & $-1$ & $0$ & $0$&$1$ & $-1$ & $0$ \\ \hline 
\end{tabular}}&\nonumber
\eea

The local, and integrated, Ward identities \equ{eqn:CSlocward}, 
\equ{eqn:WardZZ}, expressing the residual gauge invariance, are left 
unchanged, and also the discrete parity symmetry \equ{eqn:inversion} still 
holds for MCS theory.

The equations of motion induced by the action $\S_{MCS}$ 
\equ{mcsaction} are
\bea
k(\bar{\pa}A_u-\pa_u\bar{A})+i\a[\bar{\pa}(\pa\bar{A}
-\bar{\pa}{A})+\pa_u(\pa_u\bar{A}-\bar{\pa}A_u)]+\bar{J}&=&0 \label{eqn:MCSeqI} \\
k(\pa_uA-\pa A_u )
+i\a[\pa(\bar{\pa}A-\pa\bar{A}
)+\pa_u(\pa_uA-\pa A_u)]+J&=&0 \label{eqn:MCSeqII} \\
k(\pa\bar{A}-\bar{\pa}A)+i\a[2\pa\bar{\pa}A_u
-\pa_u(\bar{\pa}A+\pa\bar{A})]+b+J_u&=&0  \label{eqn:MCSeqIII} \\
A_u+J_b&=&0\ . \label{eqn:MCSeqIV}
\eea

\subsection{The boundary}

We now introduce
the planar boundary $u=0$. The decoupling condition
imposes that the generating
functional of the connected Green functions $\WW$ can be written as
\be
\WW=\WW_++\WW_- 
\ee
and, consequently, the propagators of the theory take the form \equ{eqn:scaldec}.
% \be
% \Delta_{\Phi_{1}\Phi_{2}}(x_{1},x_{2})=
% \vev{T(\Phi_{1}(x_{1})\Phi_{2}(x_{2}))}=
% \theta_{+}\Delta_{+}(x_{1},x_{2})+\theta_{-}\Delta_{-}(x_{1},x_{2}) \ ,
% \ee
% where $x =  (z,\bar{z},u)$, and $\theta_{\pm} = \theta(\pm u_{1})\theta(\pm u_{2})$.

The presence of the boundary induces in
the equations of motion boundary terms that must respect
all the constraints of locality, linearity in the 
quantum fields, power counting,
conserved quantum numbers and analiticity in the 
parameters. The most general
broken equations of motion satisfying all the requirements are

\bea
&&k(\bar{\pa}A_u-\pa_u\bar{A})+i\a[\bar{\pa}(\pa\bar{A}
-\bar{\pa}{A})+\pa_u(\pa_u\bar{A}-\bar{\pa}A_u)]+\bar{J}=\nonumber\\
&&=\d(u)\lc\a_1^+\bar{A}_++\a_1^-\bar{A}_-+\a_2^+\lp\pa_u\bar{A}\rp_++\a_2^-\lp\pa_u\bar{A}\rp_-\rc\nonumber\\
&&+\d'(u)\lc\a_3^+\bar{A}_++\a_3^-\bar{A}_-\rc 
+\d(u)\lp\a_6^+\bar{\pa}{A_u}_++\a_6^-\bar{\pa}{A_u}_-\rp\label{eqn:MCSbrokeqI}
\eea
\bea
&&k(\pa_uA-\pa A_u )
+i\a[\pa(\bar{\pa}A-\pa\bar{A})+\pa_u(\pa_uA-\pa A_u)]+J=\nonumber\\
&&=\d(u)\lc \g_1^+A_++\g_1^-A_-+\g_2^+\lp\pa_uA\rp_++\g_2^-\lp\pa_uA\rp_-\rc\nonumber\\
&&+\d'(u)\lc\g_3^+A_++\g_3^-A_-\rc+\d(u)\lp\g_4^+\pa {A_u}_++\g_4^-\pa{A_u}_-\rp\label{eqn:MCSbrokeqII}
\eea
\bea
&&k(\pa\bar{A}-\bar{\pa}A)+i\a[2\pa\bar{\pa}A_u
-\pa_u(\bar{\pa}A+\pa\bar{A})]+b+J_u=\ \ \ \ \  \nonumber\\
&&=\d(u)\lc\a_4^+\bar{\pa}A_++\a_4^-\bar{\pa}A_-+\a_5^+\pa\bar{A}_++\a_5^-\pa\bar{A}_-\rc\nonumber\\
&&+\d(u)\lp\a_8^+{A_u}_++\a_8^-{A_u}_-\rp + \d(u)\lc\a_9^+\lp\pa_uA_u\rp_++\a_9^-\lp\pa_uA_u\rp_-\rc\nonumber\\
&&+\d'(u)\lp\a_{10}^+{A_u}_++\a_{10}^-{A_u}_-\rp 
 \label{eqn:MCSbrokeqIII} 
\eea
\bea
&&A_u+J_b=0 \ .
\ \ \ \ \ \ \ \ \ \ \ \ \ \ \ \ \ \ \ \  
\ \ \ \ \ \ \ \ \ \ \ \ \ \ \ \ \ \ \ \ 
\ \ \ \ \ \ \ \ \ \ \ \ \ \ \ \ \ \ \ \ 
\label{eqn:MCSbrokeqIV}
\eea
In the above equations, $\a_i^\pm$ and $\g_i^\pm$
are constant parameters. Notice that the lower dimension of the gauge field, 
with respect to the CS case, 
allows the presence of differentiated 
delta functions $\d'(u)$ in the right-hand sides of the broken
equations of motion.

Again, we have used
the boundary fields on the left, respectively  on the right
of the boundary:
\bea
\bar{A}_\pm(Z)&=&\lim_{u\rightarrow0^\pm}{\bar{A}(x)}\\
{A}_\pm(Z)&=&\lim_{u\rightarrow0^\pm}{A(x)}\\
{A_u}_\pm(Z)&=&\lim_{u\rightarrow0^\pm}{A_u(x)}\\
\lp\pa_u\bar{A}\rp_\pm(Z)&=&\lim_{u\rightarrow0^\pm}\pa_u\bar{A}(x)\\
\lp\pa_uA\rp_\pm(Z)&=&\lim_{u\rightarrow0^\pm}\pa_uA(x)\\
\lp\pa_uA_u\rp_\pm(Z)&=&\lim_{u\rightarrow0^\pm}\pa_uA_u(x)\ ,
\eea
where $Z = (z,\bar{z})$. 

Under the parity symmetry \equ{eqn:inversion}, the boundary fields
transform according to
\be
\ba{rcl}
A_\pm&\leftrightarrow&\bar{A}_\mp\\
{A_u}_\pm&\rightarrow&-{A_u}_\mp\\
\lp\pa_uA\rp_\pm&\leftrightarrow&-\lp\pa_u\bar{A}\rp_\mp\\
\lp\pa_uA_u\rp_\pm&\leftrightarrow&\lp\pa_uA_u\rp_\mp \ .\ea
\ee

We now proceed by imposing the constraints that we 
discussed in the previous Section. 

The parity constraint yields
the following conditions:
\bea
\a_1^\pm&=&\g_1^\mp\\
\a_2^\pm&=&-\g_2^\mp\\
\a_3^\pm&=&-\g_3^\mp\\
\a_6^\pm&=&-\g_4^\mp\\
\a_4^\pm&=&-\a_5^\mp\\
\a_8^+&=&\a_8^-\\
\a_9^+&=&-\a_9^-\\
\a_{10}^+&=&-\a_{10}^- \ .
\eea

We now consider the algebraic constraint of ``compatibility''
between equations of motion. We already said in the previous Section 
that this simply corresponds to requiring that the equations of 
motion, even if broken by boundary terms, commute with each other.

After a little algebra, and recalling some basic properties of the 
$\d$-functions \cite{Schwartz}, we get
\bea
&&\a_1^+=\a_1^- = \a_1\\
&&\a_2^\pm=\a_3^\mp\\
&&\a_6^+=-\a_4^+\\
&&\a_6^-=\a_5^+\\
&&\a_9^+=-\a_{10}^+ \ .
\eea

The broken equations of motion thus become
\bea
&&k(\bar{\pa}A_u-\pa_u\bar{A})+i\a[\bar{\pa}(\pa\bar{A}
-\bar{\pa}{A})+\pa_u(\pa_u\bar{A}-\bar{\pa}A_u)]+\bar{J}=\nonumber\\
&&=\d(u)\lc\a_1\lp\bar{A}_++\bar{A}_-\rp+\a_3^-\lp\pa_u\bar{A}\rp_++\a_3^+\lp\pa_u\bar{A}\rp_-\rc\nonumber\\
&&+\d'(u)\lc\a_3^+\bar{A}_++\a_3^-\bar{A}_-\rc 
-\d(u)\lp\a_4^+\bar{\pa}{A_u}_+-\a_5^+\bar{\pa}{A_u}_-\rp\label{eqn:MCS2brokeqI}
\eea
\bea
&&k(\pa_uA-\pa A_u )
+i\a[\pa(\bar{\pa}A-\pa\bar{A})+\pa_u(\pa_uA-\pa A_u)]+J=\nonumber\\
&&=\d(u)\lc\a_1\lp A_++A_-\rp-\a_3^+\lp\pa_uA\rp_+-\a_3^-\lp\pa_uA\rp_-\rc\nonumber\\
&&-\d'(u)\lc\a_3^-A_++\a_3^+A_-\rc
-\d(u)\lp\a_5^+\pa{A_u}_+-\a_4^+\pa{A_u}_-\rp \label{eqn:MCS2brokeqII}
\eea
\bea
&&k(\pa\bar{A}-\bar{\pa}A)+i\a[2\pa\bar{\pa}A_u
-\pa_u(\bar{\pa}A+\pa\bar{A})]+b+J_u=\ \ \ \ \  \nonumber\\
&&=\d(u)\lc\a_4^+\lp\bar{\pa}A_+-\pa\bar{A}_-\rp+\a_5^+\lp\pa\bar{A}_+-\bar{\pa}{A}_-\rp\rc \nonumber\\
&&+\d(u)\lc\a_8^+\lp{A_u}_++{A_u}_-\rp-\a_{10}^+\lc\lp\pa_uA_u\rp_+-\lp\pa_uA_u\rp_-\rc\rc\nonumber\\
&&+\d'(u)\a_{10}^+\lp {A_u}_+-{A_u}_-\rp 
 \label{eqn:MCS2brokeqIII} 
\eea
\bea
&&A_u+J_b=0 \ .
\ \ \ \ \ \ \ \ \ \ \ \ \ \ \ \ \ \ \ \  
\ \ \ \ \ \ \ \ \ \ \ \ \ \ \ \ \ \ \ \ 
\ \ \ \ \ \ \ \ \ \ \ \ \ \ \ \ \ \ \ \ 
\label{eqn:MCS2brokeqIV}
\eea
Correspondingly, the local Ward identity \equ{eqn:CSlocward} acquires a boundary breaking:
\bea
&&\pa\bar{J}+\bar{\pa} J+\pa_uJ_u+\pa_ub=\d(u)\lc\a_1\lp\pa\bar{A}_++\pa\bar{A}_-
+\bar{\pa}A_++\bar{\pa}A_-\rp\rc\nonumber\\
&&+\d(u)\left\{\a_3^-\lc\pa\lp\pa_u\bar{A}\rp_+-\bar{\pa}\lp\pa_uA\rp_-\rc
+\a_3^+\lc\pa\lp\pa_u\bar{A}\rp_-
-\bar{\pa}\lp\pa_uA\rp_+\rc\right\}\nonumber\\
&&-\d(u)\lc\lp\a_4^++\a_5^+\rp\pa\bar{\pa}{A_u}_+-\lp\a_4^++\a_5^+\rp\pa\bar{\pa}{A_u}_-\rc\nonumber\\
&&+\d'(u)\lc\a_3^+\lp\pa\bar{A}_+-\bar{\pa}A_-\rp+\a_3^-\lp\pa\bar{A}_--\bar{\pa}A_+\rp\rc\nonumber\\
&&+\d'(u)\lc\a_4^+\lp\bar{\pa}A_+-\pa\bar{A}_-\rp+\a_5^+\lp\pa\bar{A}_+-\bar{\pa}{A}_-\rp\rc\nonumber\\
&&+\d'(u)\lc\a_8^+\lp{A_u}_++{A_u}_-\rp-\a_{10}^+\lc \lp \pa_uA_u\rp_+-\lp\pa_uA_u\rp_-\rc\rc\nonumber\\
&&+\d''(u)\a_{10}^+\lp {A_u}_+-{A_u}_-\rp \ ,
\label{eqn:MCSlocwardII}
\eea
which, once integrated and taking into account the axial gauge
condition $A_u=0$, yields the residual Ward identity 
\bea
&&\int du \lp\pa\bar{J}+\bar{\pa}J\rp=\a_1\lp\pa\bar{A}_++\pa\bar{A}_-+\bar{\pa}A_+
+\bar{\pa}A_-\rp+\nonumber\\
&&+\a_3^-\lc\pa\lp\pa_u\bar{A}\rp_+-\bar{\pa}\lp\pa_uA\rp_-\rc
+\a_3^+\lc\pa\lp\pa_u\bar{A}\rp_--\bar{\pa}\lp\pa_uA\rp_+\rc \ . \label{eqn:MCSresward}
\eea 
In the previous Section, we have seen 
that, in CS theory with boundary, it is the integrated residual Ward 
identity \equ{eqn:MCSresward}
which implies the existence of an algebra 
of chiral conserved currents on the boundary. Our aim
will now be to study the change of the boundary conditions induced by the additional 
Maxwell term on the fields, and the consequences on the algebra: indeed, it is
not obvious that there will be still an algebra of boundary currents. 

\subsection{The boundary conditions I: the set up}

We now infer the boundary conditions on the fields, following the 
method illustrated in the previous Section for the pure CS case.

The first step is to  set the sources to zero and integrate the broken equations of motion
between the two sides of the boundary, thus getting the conditions
\bea
&&k\lp\bar{A}_--\bar{A}_+\rp+i\a\int_{-\e}^{\e}du\ \bar{\pa}\lp\pa\bar{A}-\bar{\pa}A\rp
-i\a\lc\lp\pa_u\bar{A}\rp_--\lp\pa_u\bar{A}\rp_+\rc=\nonumber\\
&&=\a_1\lp\bar{A}_++\bar{A}_-\rp+\a_3^-\lp\pa_u\bar{A}\rp_++
\a_3^+\lp\pa_u\bar{A}\rp_-\label{eqn:MCSI1}\\
\nonumber\\
&&k\lp A_+-A_-\rp+i\a\int_{-\e}^{\e}du\ \pa\lp\bar{\pa}A-\pa\bar{A}\rp
-i\a\lc\lp\pa_uA\rp_--\lp\pa_uA\rp_+\rc=\nonumber\\
&&=\a_1\lp A_++A_-\rp-\a_3^+\lp\pa_uA\rp_+-
\a_3^-\lp\pa_uA\rp_-\label{eqn:MCSI2}\\
\nonumber\\
&&k\int_{-\e}^{\e}du\lp\pa\bar{A}-\bar{\pa}A\rp-i\a\lp\bar{\pa}A_+-\bar{\pa}A_-+\pa\bar{A}_+-\pa\bar{A}_-\rp
+\int_{-\e}^{\e}du\ b=\nonumber\\
&&\a_4^+\lp\bar{\pa}A_+-\pa\bar{A}_-\rp+\a_5^+\lp\pa\bar{A}_+-\bar{\pa}A_-\rp \ ,\label{eqn:MCSI3}
\eea 
where we have also used the axial gauge condition $A_u=0$.
Notice that \equ{eqn:MCSI2} can be obtained from the first
one by means of the parity transformation \equ{eqn:inversion}. Therefore, we can 
consider only \equ{eqn:MCSI1} and then
apply parity to the resulting conditions
to get the analogous ones generated by \equ{eqn:MCSI2}.
Taking the limit $\e\rightarrow0$ in \equ{eqn:MCSI1} and \equ{eqn:MCSI3}
we get the two conditions
\bea
&&\lp\a_1+k\rp\bar{A}_++\lp\a_3^--i\a\rp\lp\pa_u\bar{A}\rp_+=\nonumber\\
&&=-\lp\a_1-k\rp\bar{A}_--\lp\a_3^++i\a\rp\lp\pa_u\bar{A}\rp_- \label{eqn:MCSfirst}\\
\nonumber\\
&&\lp\a_4^++i\a\rp\pa\bar{A}_-+\lp\a_5^++i\a\rp\bar{\pa}A_-=\nonumber\\
&&=\lp\a_4^++i\a\rp\bar{\pa}A_++\lp\a_5^++i\a\rp\pa\bar{A}_+ \ . \label{eqn:MCSsecond}
\eea
Let us focus on \equ{eqn:MCSsecond} first. As usual, separability splits it
into
\bea
\lp\a_4^++i\a\rp\bar{\pa}A_++\lp\a_5^++i\a\rp\pa\bar{A}_+&=&0 \label{eqn:MCScons+}\\
\lp\a_4^++i\a\rp\pa\bar{A}_-+\lp\a_5^++i\a\rp\bar{\pa}A_-&=&0 \ . \label{eqn:MCScons-}
\eea
Since the parameters $\a_4^+$ and $\a_5^+$ do not appear in the 
Ward identity \equ{eqn:MCSresward}, the boundary algebra does not 
depend on them, and hence we can solve \equ{eqn:MCScons+} and 
\equ{eqn:MCScons-} by choosing~:
\be
\a_4^+=\a_5^+\neq-i\a \ , \label{eqn:MCSconscond}
\ee 
so that \equ{eqn:MCScons+} and \equ{eqn:MCScons-} yield
\bea
\bar{\pa}A_++\pa\bar{A}_+&=&0\\
\bar{\pa}A_-+\pa\bar{A}_-&=&0 \ ,
\eea
which, as in the pure CS case, can be interpreted 
as conservation relations for currents
living on the right, respectively on the left of the boundary.

On the other hand, the request for separability 
splits \equ{eqn:MCSfirst} into
\bea
\lp\a_1+k\rp\bar{A}_++\lp\a_3^--i\a\rp\lp\pa_u\bar{A}\rp_+&=&0\\
\lp\a_1-k\rp\bar{A}_-+\lp\a_3^++i\a\rp\lp\pa_u\bar{A}\rp_-&=&0 \ .
\eea
Using the parity transformation \equ{eqn:inversion} on these conditions
we get the analogous ones generated by \equ{eqn:MCSI2}:
\bea
\lp\a_1+k\rp A_--\lp\a_3^--i\a\rp\lp\pa_uA\rp_-&=&0\\
\lp\a_1-k\rp A_+-\lp\a_3^++i\a\rp\lp\pa_uA\rp_+&=&0 \ .
\eea
\newline

Since the broken equations of motion contain second-order 
derivatives, the second step towards the determination of the 
boundary conditions on the quantum fields, consists into integrating them twice. 

Let us set the sources to zero and then integrate \equ{eqn:MCS2brokeqI}:
\bea
&&\int_{-\infty}^{u}du'\ \left\{k(\bar{\pa}A_u-\pa_u\bar{A})+i\a[\bar{\pa}(\pa\bar{A}
-\bar{\pa}{A})+\pa_u(\pa_u\bar{A}-\bar{\pa}A_u)]+\bar{J}\right\}=\nonumber\\
&&=\int_{-\infty}^{u}du'\ \d(u)\lc\a_1\lp\bar{A}_++\bar{A}_-\rp
+\a_3^-\lp\pa_u\bar{A}\rp_++\a_3^+\lp\pa_u\bar{A}\rp_-\rc\nonumber\\
&&+\int_{-\infty}^{u}du'\ \d'(u)\lp\a_3^+\bar{A}_++\a_3^-\bar{A}_-\rp \ ; \label{eqn:MCSIII}
\eea
from this we will get also the conditions generated by \equ{eqn:MCS2brokeqII}
 by means of the parity transformation
\equ{eqn:inversion}, like we have done above. From \equ{eqn:MCSIII} 
we get
\bea
&&-k\bar{A}+i\a\int_{-\infty}^{u}du'\ \bar{\pa}(\pa\bar{A}-\bar{\pa}{A})
+i\a\pa_u\bar{A}=\nonumber\\
&&=\theta(u)\lc\a_1\lp\bar{A}_++\bar{A}_-\rp+\a_3^-\lp\pa_u\bar{A}\rp_+
+\a_3^+\lp\pa_u\bar{A}\rp_-\rc\nonumber\\
&&+\d(u)\lp\a_3^+\bar{A}_++\a_3^-\bar{A}_-\rp \ , \label{eqn:MCSflag}
\eea
where we have assumed that the fields and their 
derivatives vanish at infinity. 
We then integrate \equ{eqn:MCSflag}
between
the two sides of the (infinitesimal) boundary:
\bea
&&-\int_{-\e}^{\e}du\ k\bar{A}+i\a\int_{-\e}^{\e}du\int_{-\infty}^{u}du'\ \bar{\pa}(\pa\bar{A}-\bar{\pa}{A})
+i\a\int_{-\e}^{\e}du\ \pa_u\bar{A}=\nonumber\\
&&=\int_{-\e}^{\e}du\ \theta(u)\lc\a_1\lp\bar{A}_++\bar{A}_-\rp+\a_3^-\lp\pa_u\bar{A}\rp_+
+\a_3^+\lp\pa_u\bar{A}\rp_-\rc\nonumber\\
&&+\int_{-\e}^{\e}du\ \d(u)\lp\a_3^+\bar{A}_++\a_3^-\bar{A}_-\rp \ ;
\eea
taking the limit $\e\rightarrow0$ we finally get
\be
-i\a\bar{A}_-+i\a\bar{A}_+=\a_3^+\bar{A}_++\a_3^-\bar{A}_- \ ,
\ee
which can be rearranged as
\be
-\lp i\a+\a_3^-\rp\bar{A}_-=\lp-i\a+\a_3^+\rp\bar{A}_+ \ .
\ee
The request for separability again splits this condition into two separate ones
for the opposite sides of the boundary:
\bea
\lp i\a+\a_3^-\rp\bar{A}_-&=&0 \\
\lp-i\a+\a_3^+\rp\bar{A}_+&=&0 \ ;
\eea
Applying the parity transformation \equ{eqn:inversion} to these conditions we get
those generated by \equ{eqn:MCS2brokeqII}
\bea
\lp i\a+\a_3^-\rp A_+&=&0 \\
\lp-i\a+\a_3^+\rp A_-&=&0 \ .
\eea
At the end, we have obtained a set of eight boundary conditions involving
both the fields and their first derivatives,
which must be simultaneously satisfied. 

Summarizing:
\bea
\lp\a_1+k\rp\bar{A}_++\lp\a_3^--i\a\rp\lp\pa_u\bar{A}\rp_+&=&0\label{eqn:MCSboundI}\\
\lp\a_1-k\rp\bar{A}_-+\lp\a_3^++i\a\rp\lp\pa_u\bar{A}\rp_-&=&0 \\
\lp\a_1+k\rp A_--\lp\a_3^--i\a\rp\lp\pa_uA\rp_-&=&0\\
\lp\a_1-k\rp A_+-\lp\a_3^++i\a\rp\lp\pa_uA\rp_+&=&0 \\
\lp i\a+\a_3^-\rp\bar{A}_-&=&0 \\
\lp-i\a+\a_3^+\rp\bar{A}_+&=&0 \\
\lp i\a+\a_3^-\rp A_+&=&0 \\
\lp-i\a+\a_3^+\rp A_-&=&0 \ , \label{eqn:MCSboundVIII}
\eea
which must be considered together with the conservation
condition \equ{eqn:MCSconscond}:
\bea
\a_4^+=\a_5^+&\neq&-i\a\\
\bar{\pa}A_++\pa\bar{A}_+&=&0 \label{eqn:MCSsumcons+}\\
\bar{\pa}A_-+\pa\bar{A}_-&=&0 \ . \label{eqn:MCSsumcons-}
\eea
We are now left with the task of 
solving the above constraints, thus identifying all the possible
choices for the parameters and their consequences on the boundary
fields and the residual Ward identity.

\subsection{The boundary conditions II: solution(s) and algebra(s)}

The first natural request is to recover the CS result
in the limit $\a\rightarrow0$. This implies a dependence 
of the parameter $\a_1$ on $\a$ of the type
\be
\a_1=\pm k \lp1+2f(\a)\rp \ , \ \lim_{\a\rightarrow0}f(\a)=0 \label{eqn:MCSlimit}
\ee
but, as we will show later, the case $f(\a)\neq0$ 
is incompatible
with the conditions \equ{eqn:MCSsumcons+} and \equ{eqn:MCSsumcons-}
expressing the conservation condition. 
Therefore, the choice
\be
\a_1=\pm k
\ee
is compulsory. Moreover, the case $\a_1=-k$ can be obtained from
 $\a_1=k$ by means of the changes
\bea
\bar{A}_\pm&\rightarrow&\bar{A}_\mp \\
A_\pm&\rightarrow&A_\mp\\
\lp \pa_u \bar{A} \rp_\pm&\rightarrow&\lp\pa_u \bar{A} \rp_\mp\\
\lp \pa_u A \rp_\pm&\rightarrow&\lp\pa_u A \rp_\mp\\
k&\rightarrow&-k\\
\a_3^\pm&\rightarrow&-\a_3^\mp \ , \label{eqn:MCSidentif}
\eea
as it can be directly checked by comparison 
with \equ{eqn:MCSboundI}-\equ{eqn:MCSboundVIII}. Therefore,
we can focus on the eight cases with $\a_1=k$.
Reminding that the boundary algebra is determined by
the residual Ward identity 
\equ{eqn:MCSresward}, for each solution
we write the boundary conditions on the fields and the corresponding Ward 
identity:

\begin{enumerate}
\item  \label{MCSIcase}
$$
\left\{ \ba{rcl}
\a_1&=&k\\
\a_3^+&=&-i\a\\
\a_3^-&=&+i\a\\ \ea \right.
\Rightarrow A_\pm=\bar{A}_\pm=0 \nonumber
$$ 
\bea
\int du\lp\pa{\bar{J}}+\bar{\pa}J\rp&=&i\a\pa\lc\lp\pa_u\bar{A}\rp_+-\lp\pa_u\bar{A}\rp_-\rc\nonumber\\
&+&i\a\bar{\pa}\lc\lp\pa_uA\rp_+-\lp\pa_uA\rp_-\rc \ ;
\eea

\item  \label{MCSIIcase}
$$
\left\{ \ba{rcl}
\a_1&=&k\\
\a_3^+&=&-i\a\\
\a_3^-&=&-i\a\\ \ea \right.
\Rightarrow
\left \{ \ba{rcl}
\bar{A}_+&=0=&A_-\\
\lp\pa_u\bar{A}\rp_+&=0=&\lp\pa_uA\rp_- \ea \right.
$$ 
\be
\int du\lp\pa{\bar{J}}+\bar{\pa}J\rp=
k\lp\pa\bar{A}_-+\bar{\pa}A_+\rp
+i\a\lc\bar{\pa}\lp\pa_uA\rp_+-\pa\lp\pa_u\bar{A}\rp_-\rc \ ;
\ee

\item \label{MCSIIIcase}
$$
\left\{ \ba{rcl}
\a_1&=&k\\
\a_3^+&=&+i\a\\
\a_3^-&=&-i\a\\
\ea \right.
\Rightarrow \left\{ \ba{rcl}
\lp\pa_uA\rp_+=\lp\pa_u\bar{A}\rp_-&=&0\\
k\bar{A}_+-i\a\lp\pa_u\bar{A}\rp_+&=&0\\
kA_-+i\a\lp\pa_uA\rp_-&=&0
\ea \right.
$$
\bea
\int du\lp\pa{\bar{J}}+\bar{\pa}J\rp=k\lp\pa\bar{A}_-+\bar{\pa}A_+\rp \ ;
\eea

\item  \label{MCSIVcase}
$$
\left\{ \ba{rcl}
\a_1&=&k\\
\a_3^+&=&-i\a\\
\a_3^-&\neq&\mp i\a\\ \ea \right.
\Rightarrow
\left \{ \ba{rcl}
\bar{A}_\pm&=0=&A_\pm\\
\lp\pa_u\bar{A}\rp_+&=0=&\lp\pa_uA\rp_- \ea \right.
$$
\be
\int du\lp\pa{\bar{J}}+\bar{\pa}J\rp=i\a\lc\bar{\pa}\lp\pa_uA\rp_+-\pa\lp\pa_u\bar{A}\rp_-\rc \ ;
\ee

\item  \label{MCSVcase}
$$
\left\{ \ba{rcl}
\a_1&=&k\\
\a_3^+&\neq&-i\a\\
\a_3^-&=&+i\a\\ \ea \right.
\Rightarrow
\left \{ \ba{rcl}
\bar{A}_\pm&=0=&A_\pm\\
\lp\pa_uA\rp_+&=0=&\lp\pa_u\bar{A}\rp_- \ea \right.
$$
\be
\int du\lp\pa{\bar{J}}+\bar{\pa}J\rp=i\a\lc\pa\lp\pa_u\bar{A}\rp_+-\bar{\pa}\lp\pa_uA\rp_-\rc \ ;
\ee

\item \label{MCSVIcase}
$$
\left\{ \ba{rcl}
\a_1&=&k\\
\a_3^+&=&+i\a\\
\a_3^-&\neq&\mp i\a
\ea \right.
\Rightarrow 
\left\{ \ba{rcl}
A_+=\bar{A}_-=\lp\pa_uA\rp_+=\lp\pa_u\bar{A}\rp_-&=0\\
2k\bar{A}_++\lp-i\a+\a_3^-\rp\lp\pa_u\bar{A}\rp_+&=0\\
2kA_--\lp-i\a+\a_3^-\rp\lp\pa_uA\rp_-&=0
\ea \right.
$$
\be
\int du\lp\pa{\bar{J}}+\bar{\pa}J\rp=k\f{i\a+\a_3^-}{i\a-\a_3^-}\lp\pa\bar{A}_++\bar{\pa}A_-\rp \ ;
\ee

\item  \label{MCSVIIcase}
$$
\left\{ \ba{rcl}
\a_1&=&k\\
\a_3^+&\neq&\mp i\a\\
\a_3^-&=&-i\a
\ea \right.
\Rightarrow
\left\{ \ba{rcl}
\bar{A}_+=A_-&=&0\\
\lp\pa_u\bar{A}\rp_\pm=\lp\pa_uA\rp_\pm&=&0
\ea \right.
$$
\be
\int du\lp\pa{\bar{J}}+\bar{\pa}J\rp=k\lp\pa\bar{A}_-+\bar{\pa}A_+\rp \ ;
\ee

\item  \label{MCSVIIIcase}
$$
\left\{ \ba{rcl}
\a_1&=&k\\
\a_3^+&\neq&\mp i\a\\
\a_3^-&\neq&\mp i\a
\ea \right.
\Rightarrow
\left\{ \ba{rcl}
A_\pm=\bar{A}_\pm&=&0\\
\lp\pa_uA\rp_\pm=\lp\pa_u\bar{A}\rp_\pm&=&0
\ea \right.
$$
\be
\int du\lp\pa{\bar{J}}+\bar{\pa}J\rp=0 \ .
\ee

\end{enumerate}

However, not all these solutions of the system
\equ{eqn:MCSboundI}-\equ{eqn:MCSboundVIII} are acceptable. 
Indeed, we must still keep in mind that taking the limit $\a\rightarrow0$
we must recover the result found for the CS theory. 

This condition eliminates the 
\ref{MCSIcase}., \ref{MCSIVcase}., \ref{MCSVcase}.
and \ref{MCSVIIIcase}.
 Furthermore, the conservation relations
\equ{eqn:MCSsumcons+} and \equ{eqn:MCSsumcons-}, 
together with the boundary conditions, 
imply that the Ward identity for the cases
\ref{MCSVIcase}.) and \ref{MCSVIIcase}.)
becomes
\be
\int du\lp\pa{\bar{J}}+\bar{\pa}J\rp=0 \ ,
\ee
while for the case \ref{MCSIIcase}.) it takes the form
\be
\int du\lp\pa{\bar{J}}+\bar{\pa}J\rp=
+i\a\lc\bar{\pa}\lp\pa_uA\rp_+-\pa\lp\pa_u\bar{A}\rp_-\rc \ .
\ee
Recalling the form \equ{eqn:CSwardboundint} ofthe Ward identity for 
pure CS theory with boundary,  these cases do not have the correct limit $\a\rightarrow0$ either, 
and therefore they are forbidden as well.

At the end, we are left only with the case \ref{MCSIIIcase}.), which 
corresponds to \emph{Neumann} and \emph{Robin} boundary conditions
\bea
&&\left\{ \ba{rcl}
\a_1&=&k\\
\a_3^+&=&+i\a\\
\a_3^-&=&-i\a\\
\a_4^+&=&\a_5^+\neq-ik
\ea \right.
\eea

which yields on the rhs of the boundary
\bea
&&\left\{ \ba{rcl}
\lp\pa_uA\rp_+&=&0\\
k\bar{A}_+-i\a\lp\pa_u\bar{A}\rp_+&=&0\\
\bar{\pa}A_++\pa\bar{A}_+&=&0\\
\ea \right.\\
&&\int du\lp\pa{\bar{J}}+\bar{\pa}J\rp=k\lp\pa\bar{A}_-+\bar{\pa}A_+\rp \ . \label{eqn:MCSwarddef}
\eea
The Ward identity \equ{eqn:MCSwarddef} is the same as in the CS
case, and it implies both the chirality
\be
\bar{\pa}A_+=0\\ \label{eqn:MCSchirality}
\ee
and the boundary Ka\v{c}-Moody algebra
\be
\lc A_+(z),A_+(z')\rc=\f{1}{k}\pa\d(z-z')\\ \label{eqn:MCScommutation}
\ee
involving the conserved chiral current $A_+(z)$.

The corresponding solution on the lhs of the boundary can be obtained 
by parity, and read

\be\left\{ \ba{rcl}                                                                      %
\lp\pa_u\bar{A}\rp_-&=&0\\                                               %                                        %
   kA_-+i\a\lp\pa_uA\rp_-&=&0\\                                                                                                                 %
  \bar{\pa}A_-+\pa\bar{A}_-&=&0                                                           %
   \ea
   \right. \ee

   \be                                                                                     %
   {\pa}\bar{A}_-=0\\ \label{eqn:MCSchirality}                                               %
   \ee                                                                                     %

\be                                                                                     %
   \lc \bar{A}_-(\bar{z}),\bar{A}_-(\bar{z}')\rc=\f{1}{k}\bar{\pa}\d(\bar{z}-\bar{z}')   %
   \ee                                                                                     %

involving the conserved chiral current $\bar{A}_-(\bar{z})$.

This is a remarkable result, since we have found not only
that the boundary terms are uniquely fixed, but also that
there is still a Ka\v{c}-Moody algebra,
which is \emph{the same} as in the CS case. In other words, the Maxwell term
does not affect the boundary physics of the light-cone currents. 

This result is in agreement with the outcomes of  \cite{Park:1998yw} 
and  \cite{Dunne:1990hh}, where
 different approaches have been adopted, 
and a disk instead of a plane has been considered.

More recent approaches, like \cite{Ghoshal:1993tm} just define one boundary (say on 
the right) and formulate the local boundary interaction by means of a 
lagrangian or an algebraic way. The introduction of two boundaries or 
defects/impurities can be done as in \cite{Bajnok:2004jd}.

The claim of  \cite{Park:1998yw} and  \cite{Dunne:1990hh} is that
the boundary is characterized by Ka\v{c}-Moody algebras of conserved
charges with the same structure
of the CS model, thus concluding that the algebra is a consequence of the CS term
rather than of the nature of the entire model. We refer to our conclusions for a more
precise comparison.

We conclude this Chapter by illustrating the reason why the parameter $\a_1$ must be set 
to the value $k$. As we previously pointed out, the request
that the limit $\a\rightarrow0$ lead to the CS
result implies \equ{eqn:MCSlimit}. In such a case, 
it can be checked that the only
acceptable solution of the system \equ{eqn:MCSboundI}-\equ{eqn:MCSboundVIII}
is
$$
\left\{ \ba{rcl}
\a_1&=&k\lp 1+2f(\a)\rp\\
\a_3^+&=&+i\a\\
\a_3^-&=&-i\a\\
\a_4^+&=&\a_5^+\neq-i\a  \ea \right.
\Rightarrow
\left\{ \ba{rcl}
k\lp1+f(\a)\rp\bar{A}_+-i\a\lp\pa_u\bar{A}\rp_+&=&0\\
k\lp1+f(\a)\rp{A}_-+i\a\lp\pa_u{A}\rp_-&=&0\\
kf(\a)A_+-i\a\lp\pa_uA\rp_+&=&0\\
kf(\a)\bar{A}_-+i\a\lp\pa_u\bar{A}\rp_-&=&0
 \ ,
\ea \right.
$$
which corresponds to the broken residual Ward identity
\be
\int du \lp\pa\bar{J}+\bar{\pa}J\rp=
kf(\a)\lp\pa\bar{A}_++\bar{\pa}A_-\rp+k\lp1+f(\a)\rp\lp\pa\bar{A}_-+\bar{\pa}A_+\rp \ .
\eqn{eqn:MCSfinalresward}
Equation \equ{eqn:MCSfinalresward}, with the external sources set to zero, together
with separability, implies
the relations
\bea
kf(\a)\pa\bar{A}_++k\lp1+f(\a)\rp\bar{\pa}A_+&=&0\\
kf(\a)\bar{\pa}{A}_-+k\lp1+f(\a)\rp{\pa}\bar{A}_-&=&0\ ,
\eea
which are incompatible with the existence
of conserved quantities as expressed by
 \equ{eqn:MCSsumcons+} and \equ{eqn:MCSsumcons-}
unless $f(\a)=0$. Notice that this argument shows also 
that the request that the boundary currents be
conserved automatically implies their chirality.

\section{Conclusions}

In this paper the three 
dimensional Maxwell-Chern-Simons (MCS) theory with planar boundary has been 
considered, following 
the pioneering work of Symanzik  \cite{Symanzik:1981wd}.

We adopted a method similar to the one introduced in  \cite{Emery:1991tf}, which 
avoids the problem of regularizing the boundary action, which is 
necessarily $\d$-function dependent, but we modified it in a significant manner. 
In fact, one of the main steps towards the realization of the Symanzik's 
constraint of separability, which implements the introduction of a 
boundary in quantum field theory, is the direct computation of the 
propagators of the theory, taking into account, of course, the 
boundary interactions. While this is quite feasible in CS theory, 
in other more complicated, and physically more relevant cases -- of 
which the MCS model is an example -- this could be a much more 
difficult task. In fact, we have been able to reach the same 
goals, as in the  simpler case of pure CS theory, without a direct computation of 
the propagators of the theory.

As already pointed out, we stress that  
the addition of the Maxwell term to the CS action spoils its topological 
character, and thus it is not at all granted that the same results 
can be obtained.

Indeed, we found that, under certain conditions,
{\bf
\begin{enumerate}
    \item chiral conserved currents living on the boundary exist,
    \item which satisfy, like in the CS case, a Ka\v{c}-Moody 
    algebra, whose central charge coincides with the inverse 
    of the CS coupling constant.
\end{enumerate}
}
In particular, we have found that there are many
possible boundary terms compatible with Symanzik's
constraint of separability, each associated
with different boundary conditions on the fields. 
However, the requirement to recover the CS case in the limit
of vanishing Maxwell term, uniquely sets the boundary parameters
to values which imply \emph{Neumann and Robin boundary conditions}
on the fields on both sides of the boundary. In this
situation, the same residual Ward identity 
of the CS case holds, thus implying the results 1. and 2. above 
mentioned.

This result leads us to conclude that the boundary 
physics is independent
of the Maxwell term which breaks the topological character
of the CS model, even though the Maxwell boundary contributions to the 
equations of motion are nontrivial. This is a remarkable result, 
since it 
suggests that the existence of a boundary Ka\v{c}-Moody algebra
of conserved chiral currents
depends only on the presence of a CS term in the theory, rather
than on the particular theory itself.

We point out that our main results hold also for the nonabelian
model, namely the Yang-Mills-Chern-Simons action. Indeed, the 
conservation of boundary currents, their chirality and the central charge of 
their Ka\v{c}-Moody algebra are entirely determined by the quadratic part of 
the action only, and therefore are shared with the nonabelian 
extension, which affects the vertices of the theory.

Our result, concerning the existence of a Ka\v{c}-Moody
boundary algebra with the same structure
of the CS model, somehow agrees 
with Deser's recent statement that, for spin one and 
in three spacetime dimensions, 
\emph{``everything is CS''} \cite{Deser:2008rm}. 
In Ref.  \cite{Deser:2008rm}, in fact,
he shows how a 
Yang-Mills-Chern-Simons (YMCS)
action can be rewritten
in the form of a pure CS action in terms of a new
field. 

But,
even in the earlier work  \cite{Lemes:1998md},
three dimensional Yang-Mills gauge theories in the presence of 
the Chern-Simons action were shown to be generated by the pure 
topological Chern-Simons term through nonlinear covariant 
redefinitions of the gauge field. And indeed these claims are 
supported by our results, which do not depend on the presence of a 
bulk Maxwell term.

Another point of contact is with the outcome
of  \cite{Andrade:2005ur} for the (2+1) dimensional
black hole coupled with electrodynamics,
where it is shown that the black hole
coupled to the pure CS theory does not
 change configuration
when a Maxwell term is \emph{``turned on''}.

In the framework of the Fractional Quantum Hall Effect,
our result justifies the fact 
that an additional Maxwell
term to the pure CS effective low energy
model describes the gap of the elementary bulk 
excitations, without affecting the properties of the edge states
 \cite{Wenbook}.

Finally, we are in agreement also 
with 
 \cite{Park:1998yw} and
  \cite{Dunne:1990hh},              
where the 
nonabelian YMCS theory
defined on a manifold with boundary is considered in
the Weyl gauge.
In Ref.  \cite{Park:1998yw}, 
a  (2+1) dimensional disk is considered, 
and the Dirac's procedure is followed. The boundary
Ka\v{c}-Moody algebra is then obtained
as the projection of the
Dirac's brackets on the boundary of the disk.

In  \cite{Dunne:1990hh}
the discussion is instead carried out in terms of the Gauss
law generators, and the algebra in terms of the fields 
is just
a consequence. Nevertheless, they both 
find that also in the YMCS
theory there is a boundary algebra, which 
is the same as that
of the CS model.

However, our approach is not only simpler, but also
 stays in a more
general framework.
In fact, in both  \cite{Park:1998yw} and  \cite{Dunne:1990hh} the boundary 
conditions are chosen \emph{a priori}, 
provided that they satify some general requirements, 
while we find all possible boundary conditions compatible
with the simple and fundamental requests of locality
and separability, without requiring any other constraint.
%Further, in  \cite{Park:1998yw} and  \cite{Dunne:1990hh} the boundary is described by a 
%boundary action which is completely
%determined by the variation of the bulk
%action under a residual gauge transformation.

Moreover, in our description the boundary 
%- and hence the 
%Ward identity for the residual
%gauge invariance on the boundary - 
is the most general one 
compatible with very general 
principles - like locality, power counting and helicity conservation - 
and with the algebraic structure of the theory.

We conclude our discussion by indicating other possible 
further developments. 

In fact, the three dimensional CS theory is not the 
only possible topological field theory. The other important Schwarz type 
topological field theory  \cite{Witten:1988ze} 
is represented by the BF models 
 \cite{Deser}, which, 
contrarily to the CS theory, which is intrinsically three dimensional, 
exist in an arbitrary number of dimensions.

In three dimensions, the BF model with boundary has been studied in 
 \cite{Maggiore:1992ki}, 
where a richer algebraic structure than the CS case has been found, 
always of the  Ka\v{c}-Moody type.

A natural possible extension could therefore 
be the study of the effect of a $(d-1)$
dimensional boundary in the $d$--dimensional BF theory. The 
investigation of the possible boundary algebra in these cases is a 
challenge.


\begin{thebibliography}{999}
\bibitem{Symanzik:1981wd} 
  K.~Symanzik, 
  %\emph{Schrodinger Representation 
  %and Casimir Effect In Renormalizable Quantum
  %Field Theory}, 
  Nucl.\ Phys.\  B {\bf 190} (1981) 1.
%                    
\bibitem{Witten:1988hf}
  E.~Witten,
  %``Quantum field theory and the Jones polynomial,''
  Commun.\ Math.\ Phys.\  {\bf 121}, 351 (1989).
  %%CITATION = CMPHA,121,351;%%
%
\bibitem{Moore:1989yh}
  G.~W.Moore and N.~Seiberg,
  %``Taming the Conformal Zoo,''
  Phys.\ Lett.\  B {\bf 220}, 422 (1989).
  %%CITATION = PHLTA,B220,422;%%
%
\bibitem{Casimir:1948dh}
  H.B.G.~Casimir,
  %``On the Attraction Between Two Perfectly Conducting Plates,''
  Indag.\ Math.\  {\bf 10}, 261 (1948).
%
\bibitem{Wenbook}   
  X.-G.~Wen,
  %``Theory of the edge states in fractional quantum Hall effects,'' 
  Int.\ J.\ Mod.\ Phys.\  B {\bf 6}, 1711 (1992);\\
  \emph{Quantum Field Theory of Many Body Systems: from the Origin
  of Sound to an Origin of Light and Electrons (Oxford Graduate Texts)} 
  (Oxford University Press, USA, 2007). 
%
\bibitem{Banados:1992wn}
  M.~Ba\~{n}ados, C.~Teitelboim and J.~Zanelli,
  %``The Black hole in three-dimensional space-time,''
  Phys.\ Rev.\ Lett.\  {\bf 69}, 1849 (1992);
%\bibitem{Banados:1992gq}
  M.~Ba\~{n}ados, M.~Henneaux, C.~Teitelboim and J.~Zanelli,
  %``Geometry of the (2+1) black hole,''
  Phys.\ Rev.\  D {\bf 48}, 1506 (1993).
%
\bibitem{Diehl}
  H.W.~Diehl, 
  \emph{Field Theoretical Approach to the Critical Behaviour
  at Surface}, in \emph{Phase Transitions and Critical Phenomena Vol. 8},
  (edited by C.~Domb and J.L.~Lebowitz, Academic Press, 1986).
%
\bibitem{Witten:1988ze}
  E.~Witten,
  %``Topological Quantum Field Theory,''
  Commun.\ Math.\ Phys.\  {\bf 117}, 353 (1988).
%
\bibitem{Birmingham:1991ty}
  D.~Birmingham, M.~Blau, M.~Rakowski and G.~Thompson,
  %``Topological field theory,''
  Phys.\ Rept.\  {\bf 209}, 129 (1991).
  %%CITATION = PRPLC,209,129;%%
%
\bibitem{Kac:1967jr}
  V.G.~Ka\v{c},
  %``Simple Graded Algebras Of Finite Growth,''
  Funct.\ Anal.\ Appl.\  {\bf 1}, 328 (1967);\\
%\bibitem{Moody:1966gf}
  R.V.~Moody,
  %``Lie Algebras associated with generalized Cartan matrices,''
  Bull.\ Am.\ Math.\ Soc.\  {\bf 73}, 217 (1967).
%
\bibitem{Witten:1983ar}
  E.~Witten,
  %``Nonabelian bosonization in two dimensions,''
  Commun.\ Math.\ Phys.\  {\bf 92}, 455 (1984).
%
\bibitem{Belavin:1984vu}
  A.~A.~Belavin, A.~M.~Polyakov and A.~B.~Zamolodchikov,
  %``Infinite conformal symmetry in two-dimensional quantum field theory,''
  Nucl.\ Phys.\  B {\bf 241}, 333 (1984).
%
\bibitem{Deser}
  S.~Deser, 
  Phys.\ Rev.\ {\bf 178}, 1931 (1969);\\
  D.Z.~Freedman and P.K.~Townsend,
  Nucl.\ Phys.\ B {\bf 177} 282.
%
\bibitem{Maggiore:1992ki}
  N.~Maggiore and P.~Provero,
  %``Chiral Current Algebras In Three-Dimensional Bf Theory With Boundary,''
  Helv.\ Phys.\ Acta {\bf 65}, 993 (1992).
% 
\bibitem{Witten:1988hc}
  E.~Witten,
  %``(2+1)-Dimensional Gravity as an Exactly Soluble System,''
  Nucl.\ Phys.\  B {\bf 311}, 46 (1988).
%
\bibitem{Deser:1983dr}
  S.~Deser and R.~Jackiw,
  %``Three-Dimensional Cosmological Gravity: Dynamics Of Constant Curvature,''
  Annals Phys.\  {\bf 153}, 405 (1984);
%\bibitem{Witten:1989ip}
  E.~Witten,
  %``QUANTIZATION OF CHERN-SIMONS GAUGE THEORY WITH COMPLEX GAUGE GROUP,''
  Commun.\ Math.\ Phys.\  {\bf 137}, 29 (1991).
%
\bibitem{Banados:1998ta}
  M.~Banados, T.~Brotz and M.~E.~Ortiz,
  %``Boundary dynamics and the statistical mechanics of the 2+1 dimensional
  %black hole,''
  Nucl.\ Phys.\  B {\bf 545}, 340 (1999);
%
\bibitem{Park:1998qk}
  M.I.~Park,
  %``Statistical entropy of three-dimensional Kerr-de Sitter space,''
  Phys.\ Lett.\  B {\bf 440}, 275 (1998)
%
\bibitem{Wilczek:1990ik}
  F.~Wilczek,
  \emph{Fractional Statistics and Anyon Superconductivity}
  %\href{http://www.slac.stanford.edu/spires/find/hep/www?irn=2455790}{SPIRES entry}
  (Singapore: World Scientific, Singapore, 1990).
%
\bibitem{Deser:1981wh}
  S.~Deser, R.~Jackiw and S.~Templeton,
  %``Topologically massive gauge theories,''
  Annals Phys.\  {\bf 140}, 372 (1982).
%
\bibitem{Milton:1990yj}
  K.A.~Milton and Y.J.~Ng,
  %``The Maxwell-Chern-Simons Casimir Effect,''
  Phys.\ Rev.\  D {\bf 42}, 2875 (1990).
%
\bibitem{Andrade:2005ur}
  T.~Andrade, M.~Banados, R.~Benguria and A.~Gomberoff,
  %``The 2+1 charged black hole in topologically massive electrodynamics,''
  Phys.\ Rev.\ Lett.\  {\bf 95}, 021102 (2005).
%
\bibitem{Blasi:1990bk}
  A.~Blasi and R.~Collina,
  %``Chern-Simons model in the Landau gauge and its connection to the Ka\v{c}-Moody
  %algebra,''
  Nucl.\ Phys.\ Proc.\ Suppl.\   {\bf 18B} (1991) 16;
%\bibitem{Blasi:1990pf}
%  A.~Blasi and R.~Collina,
  %``The Chern-Simons Model With Boundary: A Cohomological Approach,''
  Int.\ J.\ Mod.\ Phys.\  A {\bf 7}, 3083 (1992).
%
\bibitem{Emery:1991tf}
  S.~Emery and O.~Piguet,
  %``Chern-Simons theory in the axial gauge: Manifold with boundary,''
  Helv.\ Phys.\ Acta {\bf 64}, 1256 (1991).
  %
\bibitem{BFMMS}
  A.~Blasi, D.~Ferraro, N.~Maggiore, N.~Magnoli and M.~Sassetti,
  Ann.\ Phys.\ {\bf 17}, 885 (2008).
  %
\bibitem{Soldati:1991book}
 A.~Bassetto, G.~Nardelli and R.~Soldati, 
 \emph{Yang-Mills Theories in Algebraic Non Covariant Gauges : Quantization and Renormalization},
 World Scientific, Singapore (1991).
%
 %
\bibitem{Park:1998yw}
  M.~I.~Park,
  %``Symmetry Algebras in Chern-Simons Theories with Boundary: Canonical Approach,''
  Nucl.\ Phys.\  B {\bf 544}, 377 (1999).
%
\bibitem{Dunne:1990hh}
  G.V.~Dunne and C.A.~Trugenberger,
  %``ODD DIMENSIONAL GAUGE THEORIES AND CURRENT ALGEBRA,''
  Annals Phys.\  {\bf 204}, 281 (1990).
%
  \bibitem{Balachandran:1993tm}
  A.P.~Balachandran, L.~Chandar, E.~Ercolessi, T.R.~Govindarajan and R.~Shankar,
  %``Maxwell-Chern-Simons electrodynamics on a disk,''
  Int.\ J.\ Mod.\ Phys.\  A {\bf 9}, 3417 (1994).
%
  \bibitem{Mermin:1979zz}
  N.~D.~Mermin,
  %``The topological theory of defects in ordered media,''
  Rev.\ Mod.\ Phys.\  {\bf 51}, 591 (1979).
%
\bibitem{defectbook}
  D.A.~Drabold and S.~Estreicher,
  \emph{Theory of Defects in Semiconductors}
  (Springer, 2007).
%
\bibitem{Berezinsky:1998ft}
  V.~Berezinsky, P.~Blasi and A.~Vilenkin,
  %``Ultra high energy gamma rays as signature of topological defects,''
  Phys.\ Rev.\  D {\bf 58}, 103515 (1998).
%
\bibitem{Aharonian:1992qf}
  F.~A.~Aharonian, P.~Bhattacharjee and D.~N.~Schramm,
  %``Photon / proton ratio as a diagnostic tool for topological defects  as the
  %sources of extremely high-energy cosmic rays,''
  Phys.\ Rev.\  D {\bf 46}, 4188 (1992).
%
  \bibitem{Mintchev:2007qt}
  M.~Mintchev and E.~Ragoucy,
  %``Algebraic approach to multiple defects on the line and application to
  %Casimir force,''
  J.\ Phys.\ A  {\bf 40}, 9515 (2007).
  %
  \bibitem{Weinberg1}
  S.~Weinberg, \emph{The Quantum Theory of Fields
  Vol.1: Foundations} (Cambridge University Press, Cambridge, 1995).
%
\bibitem{Blasi:1992mm}
  A.~Blasi, R.~Collina and J.~Sassarini,
  %``Finite Casimir effect in quantum field theory,''
  Int.\ J.\ Mod.\ Phys.\  A {\bf 9} (1994) 1677.
%
  \bibitem{Dirac:1949cp}
  P.A.M.~Dirac,
  %``Forms Of Relativistic Dynamics,''
  Rev.\ Mod.\ Phys.\  {\bf 21}, 392 (1949).
%
  \bibitem{Schwartz}
  L.~Schwartz,
  \emph{Introduction to the Theory of Distributions},
  (University of Toronto Press, Toronto, 1960).
  %
  \bibitem{Ghoshal:1993tm}
  S.~Ghoshal and A.~B.~Zamolodchikov,
  %``Boundary S matrix and boundary state in two-dimensional integrable quantum
  %field theory,''
  Int.\ J.\ Mod.\ Phys.\  A {\bf 9}, 3841 (1994)
  [Erratum-ibid.\  A {\bf 9}, 4353 (1994)]
  [arXiv:hep-th/9306002].
  %
  \bibitem{Bajnok:2004jd}
  Z.~Bajnok and A.~George,
  %``From defects to boundaries,''
  Int.\ J.\ Mod.\ Phys.\  A {\bf 21}, 1063 (2006)
  [arXiv:hep-th/0404199].
  %
  \bibitem{Deser:2008rm}
  S.~Deser,
  %``Distended Topologically Massive Electrodynamics,''
  published in D.~Grumiller, A.~Rebhan, D.~Vassilevich, 
  \emph{Fundamental Interactions},  (World Scientific, 2009), 
  e-Print: arXiv:0810.5384 [hep-th].
%
 \bibitem{Lemes:1998md}
   V.E.R.~Lemes, C.~Linhares de Jesus, S.P.~Sorella, L.C.Q.~Vilar and O.S.~Ventura,
%   ``Chern-Simons as a geometrical setup for three-dimensional gauge theories,''
   Phys.\ Rev.\  D {\bf 58}, 045010 (1998).
%   [arXiv:hep-th/9801021].
   %
\end{thebibliography}
\end{document}